\documentclass[a4paper,11pt]{article}
 
\usepackage{jheppub} 
           
\usepackage[T1]{fontenc} 
\usepackage{amsmath}  
\usepackage{amssymb}  
\usepackage{latexsym}
\usepackage{enumitem} 
\usepackage{lipsum}
\usepackage{graphicx}
\usepackage{subfig} 
\usepackage{empheq}
\usepackage{upgreek} 
\usepackage{cancel}
\usepackage{ytableau}
\usepackage{hyperref}
\usepackage{pifont}
\usepackage{ytableau}
\usepackage{blkarray}
\usepackage{multirow}
\usepackage{lscape}
\usepackage{float}
\usepackage{tikz-feynman,contour}

\usepackage{braket}
\usepackage{amssymb}
\usepackage{amsmath}
\usepackage{amsthm}
\usepackage{mathrsfs}
\usepackage{skak}
\usepackage{framed}
\usepackage{hyperref}
\usepackage{slashed}
\usepackage{array}

\hypersetup{colorlinks=true}
\hypersetup{linkcolor=violet}
\hypersetup{citecolor=violet}
\hypersetup{urlcolor=violet}

\numberwithin{equation}{section}

\usepackage
{datetime}
\usepackage{fancyhdr}
\usepackage{tikz, pgf}
\usetikzlibrary{shapes.misc}
\usetikzlibrary{shapes}
\usetikzlibrary{fit}
\usetikzlibrary{arrows}

\usetikzlibrary{decorations.pathmorphing}	
\usetikzlibrary{decorations.markings}
 \usetikzlibrary{patterns}

\tikzset{
    sugra/.style={decorate, decoration={snake}, draw=black},
    scalarphi/.style={dashed,draw=black, postaction={decorate},
        },
    scalarchi/.style={draw=brown}, 
    hwbou/.style={draw=blue, postaction={decorate}, ultra thick
        },
    vector/.style={draw=blue,decorate, decoration={snake}, draw},
	provector/.style={decorate, decoration={snake,amplitude=2.5pt}, draw},
	antivector/.style={decorate, decoration={snake,amplitude=-2.5pt}, draw},
   	 fermion/.style={draw=cyan, postaction={decorate},
        decoration={markings,mark=at position .55 with {\arrow[draw=black]{>}}}},
    fermionbar/.style={draw=cyan, postaction={decorate},
        decoration={markings,mark=at position .55 with {\arrow[draw=black]{<}}}},
    chspin/.style={draw=red, postaction={decorate},
        decoration={markings,mark=at position .55 with {\arrow[draw=black]{>}}}},
    chspinbar/.style={draw=red, postaction={decorate},
        decoration={markings,mark=at position .55 with {\arrow[draw=black]{<}}}},  
    fermionnoarrow/.style={draw=black},
    gluon/.style={decorate, draw=purple,
        decoration={coil, amplitude=4pt, segment length=5pt}},
    scalar/.style={dashed,draw=black, postaction={decorate},
        decoration={markings,mark=at position .55 with {\arrow[draw=black]{>}}}},
    scalarbar/.style={dashed,draw=black, postaction={decorate},
        decoration={markings,mark=at position .55 with {\arrow[draw=black]{<}}}},
    electron/.style={draw=black, postaction={decorate},
        decoration={markings,mark=at position .55 with {\arrow[draw=black]{>}}}},
    scalarnoarrow/.style={dashed, draw=black},
    electron/.style={draw=black, postaction={decorate},
        decoration={markings, mark=at position .55 with {\arrow[draw=black]{>}}}},
	bigvector/.style={decorate, decoration={snake, amplitude=4pt}, draw},
    photon/.style={draw=violet, decorate, decoration={snake}, draw},
    higgs/.style={dashed, draw=black, postaction={decorate},
        },	
        goldstone/.style={draw=brown, postaction={decorate},
        },    
          ghost/.style={dashed, draw=blue, postaction={decorate},
        decoration={markings, mark=at position .55 with {\arrow[draw=black]{>}}}
        },  
          antighost/.style={dashed, draw=blue, postaction={decorate},
        decoration={markings, mark=at position .55 with {\arrow[draw=black]{<}}}
        }, 
            scalartwo/.style={dashed,draw=brown, postaction={decorate},
        decoration={markings,mark=at position .55 with {\arrow[draw=black]{>}}}},
    scalarbartwo/.style={dashed,draw=brown, postaction={decorate},
        decoration={markings,mark=at position .55 with {\arrow[draw=black]{<}}}}, 
    fermiontwo/.style={draw=purple, postaction={decorate},
        decoration={markings,mark=at position .55 with {\arrow[draw=black]{>}}}},
    fermionbartwo/.style={draw=purple, postaction={decorate},
        decoration={markings,mark=at position .55 with {\arrow[draw=black]{<}}}},    
        realscalar/.style={draw=black}, 
        fakerealscalar/.style={draw=white}, 
        realscalarone/.style={ draw=black},
    	realscalartwo/.style={draw=brown},    	    pseudoscalar/.style={draw=brown},
        mgluon/.style={decorate, draw=blue,
        	decoration={coil,amplitude=4pt, segment length=5pt}},
         weylfermion/.style={draw=orange, postaction={decorate},
        decoration={markings,mark=at position .55 with {\arrow[draw=black]{>}}}},
         weylfermionbar/.style={draw=orange, postaction={decorate},
        decoration={markings,mark=at position .55 with {\arrow[draw=black]{<}}}}, 
    majorana/.style={draw=cyan, postaction={decorate},
        decoration={markings,mark=at position .55 with {\arrow[draw=black]{><}}}},
    majoranabar/.style={draw=cyan, postaction={decorate},
        decoration={markings,mark=at position .55 with {\arrow[draw=black]{><}}}},    
   	wboson/.style={draw=blue,decorate, decoration={snake,amplitude=4pt}, draw},  
    zboson/.style={draw=violet, decorate, decoration={snake}, draw},   
    lepton/.style={draw=black, postaction={decorate},
        decoration={markings, mark=at position .55 with {\arrow[draw=black]{>}}}},
    leptonbar/.style={draw=black, postaction={decorate},
        decoration={markings, mark=at position .55 with {\arrow[draw=black]{<}}}}, 
    clepton/.style={draw=cyan, postaction={decorate},
        decoration={markings, mark=at position .55 with {\arrow[draw= black]{>}}}},
    cleptonbar/.style={draw=cyan, postaction={decorate},
        decoration={markings, mark=at position .55 with {\arrow[draw=black]{<}}}},   
   nlepton/.style={draw=orange, postaction={decorate},
        decoration={markings, mark=at position .55 with {\arrow[draw=black]{>}}}},
    nleptonbar/.style={draw=orange, postaction={decorate},
        decoration={markings, mark=at position .55 with {\arrow[draw=black]{<}}}},              
        graviton/.style={draw=blue, decorate, decoration={snake, amplitude=4pt}, draw},  
        spinj/.style={draw=red, decorate, decoration={snake, amplitude=4pt}, draw},  
        bgraviton/.style={draw=blue, decorate, decoration={snake, amplitude=4pt}, draw},  
        gravitino/.style={draw=red, postaction={decorate}, 
        decoration={snake,  markings, mark=at position .55 with {\arrow[draw=black]{><}}}},
    	gravitinobar/.style={draw=red, postaction={decorate},
        decoration={snake, markings, mark=at position .55 with {\arrow[draw=black]{><}}} },  
    phir/.style={draw=blue, postaction={decorate},},
   phil/.style={dashed,draw=blue,},
     phiav/.style={draw=cyan, postaction={decorate},},
   phidif/.style={dashed,draw=cyan,},  
    chir/.style={draw=red, postaction={decorate},},
   chil/.style={dashed,draw=red,},  
}

\newcommand{\treelevelthreepointskel}[4]{
\begin{scope}[shift={(0,0)}, rotate=#4]
	\draw[#1] (0,0)--(1.5,0);    
	\draw[#2][rotate=120] (0,0)--(1.5,0);  
	\draw[#3][rotate=-120](0,0)--(1.5,0);  

\end{scope}	 
}

\newcommand{\treelevelthreepoint}[4]{
\begin{scope}[shift={(0,0)}, rotate=#4]
	\draw[#1] (0,0)--(1.5,0);   
	\draw[#2][rotate=120] (0,0)--(1.5,0);   
	\draw[#3][rotate=-120] (0,0)--(1.5,0);   
	
	\draw[->] (1.5,.5)--(.75,.5);
	\draw[->][rotate=120] (1.5,.5)--(.75,.5);
	\draw[->][rotate=-120] (1.5,-.5)--(.75,-.5);

	\node at (-.75-.5,.4) {$k_2$}; 
	\node at (-.75-.5,-.4) {$k_3$};
	\node at (1.25,.75) {$k_1$};
	
\end{scope}	 
}

\newcommand{\labeltreelevelthreepoint}[4]{
\begin{scope}[shift={(0,0)}, rotate=#4]

\node at (2,0) {#1}; 
\begin{scope}[shift={(0,0)},rotate=120]	
\node at (2,0) {#2};
\end{scope}	
\begin{scope}[shift={(0,0)},rotate=240]	
	\node at (2,0) {#3};
\end{scope}	
	
\end{scope}	 
} 

\newcommand{\treefourpointexchange}[6]{
\begin{scope}[rotate=#6]

\begin{scope}[shift={(-.75,0)} ]

\treelevelthreepointskel{fakerealscalar}{#1}{#2}{0}	
\begin{scope}[shift={(1.5,0)}]	
\treelevelthreepointskel{#5}{#3}{#4}{180}	
\end{scope}	

\end{scope}	 
\end{scope}	 
}

\newcommand{\treefourpointcrossexchange}[6]{
\begin{scope}[rotate=#6]

\begin{scope}[shift={(-.75,0)} ]

	\draw[#5] (0,0)--(1.5,0);    
	\draw[#2][rotate=-120](0,0)--(1.5,0); 

	\draw[#4][rotate=30](0,0)--(2.6,0); 
	\filldraw[white] (.75,.45) circle (.2);  

\begin{scope}[shift={(1.5,0)}, rotate=0]
	\draw[#3][rotate=-60] (0,0)--(1.5,0);
	
	\draw[#1][rotate=150](0,0)--(2.6,0);        
\end{scope}	 

\end{scope}	 
\end{scope}	 
}

\newcommand{\labeltreefourpointexchange}[6]{
\begin{scope}[rotate=#6]

\begin{scope}[shift={(-.75,0)} ]

\node at (.75,-.5) {#5}; 

\labeltreelevelthreepoint{}{#1}{#2}{0} 
\begin{scope}[shift={(1.5,0)}]	
\labeltreelevelthreepoint{}{#3}{#4}{180}
\end{scope} 
	
\end{scope}	 
\end{scope}	 
}

\usepackage{comment}
\definecolor{cobalt}{rgb}{0.0, 0.28, 0.67}

\specialcomment{rudranote}{%
  \begingroup \color{cobalt}}{%
  \endgroup}

\tikzstyle{block} = [draw, rectangle, 
    minimum height=3em, minimum width=6em]

\def\tgem{ \widetilde{\textrm{g}}_{\textrm{em}} }
 
\def\gem{ \textrm{g}_{\textrm{em}} }

 \definecolor{ao(english)}{rgb}{0.0, 0.5, 0.0}
\usepackage{pifont}

\usepackage{comment}

 \specialcomment{rudradetails}{%
  \begingroup \color{blue}}{%
  \endgroup} 

\newcommand{\spin}{\mathcal{S}}

\def\bfi{\bf{\color{red}I}}
\def\bfj{\bf{\color{cyan} J}}

\newcommand{\spp}[2]{(#1 \cdot #2)}

\def\iimg{ {\bf i}}

\def\diag{\textrm{diag}}

\hbadness=\maxdimen
\vbadness=\maxdimen

\allowdisplaybreaks[1]

\newcommand{\SHAA}[2]{\langle  #1 #2\rangle}
\newcommand{\SHBB}[2]{[  #1 #2]}
\newcommand{\SHAB}[3]{\langle  #1|#2|#3]}

\def\bfi{\bf{\color{red}I}}
\def\bfj{\bf{\color{blue} J}}

\title{\boldmath Compton amplitude and Contact term(s) in the Spinor Helicity formalism}

\vspace{10mm}

\author{Aakash Kumar,}
\author{ Arnab Rudra \&}
\author{ Rahul Shaw}
\author[]{\\ }
\vspace*{4.0ex} 
\vspace{10mm}
\affiliation[]{Indian Institute of Science Education and Research Bhopal,\\
 Bhopal Bypass Road, Bhauri, Madhya Pradesh 462066, India.\\ }
\vspace*{4.0ex} 
\vspace*{2.0ex}
\emailAdd{aakash19@iiserb.ac.in}
\emailAdd{rudra@iiserb.ac.in} 
\emailAdd{rahulshaw.phy@gmail.com}  

\abstract{In gauge theories, contact terms play an important role in ensuring gauge invariance. In the spinor-helicity formalism, the choice of a gauge-fixing condition manifests itself in the form of the choice of reference vector to write the massless polarization vector(s). However, this choice must be irrelevant in any gauge-invariant observable. We use this principle to determine contact terms for electromagnetic Compton amplitude. We considered three-point amplitude between two massive particles \& a photon to be one which is responsible for soft photon theorem/Coulomb interaction. We demonstrate that it is possible to use the above-mentioned principle to find the contact term for the tree-level Compton amplitude which makes it gauge invariant, unitary and free of spurious poles. We also verify the results with the amplitude factorization approach.}

\begin{document} 
\maketitle

\newpage
 
 \section{Introduction}

 In textbooks, the scattering of light (photons) from an electron is known as Compton scattering. In the early days of quantum theory, this was one of the experiments that confirmed the corpuscular nature of light. In the modern literature, the name Compton is used in a generalized sense: any scattering of two massless and two massive particles is named Compton scattering. 
 
In recent times, there has been a surge in the application of Compton scattering in understanding gravitational waves \cite{Abbott_2016,Abbott_2017} which are emitted from a binary merger. Quantum scattering amplitudes have turned out to be essential in describing the inspiral phase of the merger. The authors of \cite{Bjerrum-Bohr:2013bxa,Neill:2013wsa} introduced methods to use Feynman diagrams to do classical general relativity computations (see \cite{Bjerrum-Bohr:2018xdl} for reviews of this approach). In the recent literature \cite{Vines:2018gqi,  Guevara:2019fsj, Maybee:2019jus, Bautista:2021wfy, Bautista:2022wjf, Guevara:2017csg, Chung:2018kqs,Vines:2017hyw,Guevara:2018wpp, Arkani-Hamed:2019ymq, Aoude:2020onz, Ochirov:2018uyq, Chiodaroli:2021eug, Johansson:2019dnu,Arkani-Hamed:2017jhn}, it has been shown that the classical scattering of black holes can also be reproduced using the quantum Compton amplitudes in the large spin and large mass limit called as the classical limit. For example, the scattering of two Schwarzschild black holes can be reproduced from Compton scattering in a theory of a graviton and a massive scalar. One needs to consider the classical limit of the scattering to match it with general relativity. A more challenging problem is to reproduce the analogous result for Kerr black holes. This requires gravitational Compton amplitudes involving massive particles of arbitrary spin, followed by an appropriate classical limit of the resulting spinning amplitudes; see \cite{Buonanno:2022pgc} for a recent review. The main difficulty is that such amplitudes are not known in a completely systematic way for arbitrary spin. From a field-theoretic perspective, this is tied to the absence of a quantum field of finitely many massive higher-spin particles in flat spacetime. Although string theory provides a consistent higher spin Compton amplitude, it contains an infinite tower of massive higher-spin states.

 A lot of progress in computing Compton amplitudes has been made using the on-shell approaches like the double copy \cite{KAWAI19861,Bern:2010ue,Bern:2008qj,Vazquez-Holm:2025ztz} and BCFW recursion/amplitude factorization using the spinor helicity formalism \cite{Aoude:2020onz, Ochirov:2018uyq, Chiodaroli:2021eug,  Johansson:2019dnu,Arkani-Hamed:2017jhn,Chung:2018kqs}. It is known that in $3+1$ dimensions, there are $2\spin+1$ kinematically inequivalent three-point amplitudes of two massive spin $\spin$ particles and a photon/graviton \cite{Aoude:2020onz, Arkani-Hamed:2017jhn}. The authors of  \cite{Aoude:2020onz, Arkani-Hamed:2017jhn} argued that a particular linear combination of these $2\mathcal{S}+1$ couplings is special in the sense that they give rise to better behavior of the amplitude at high energies \cite{Arkani-Hamed:2017jhn}. For future purposes, we refer to this combination as UV minimal/AHH minimal \cite{Newman:1965tw, Lynden-Bell:2004coe, Scheopner:2023rzp, Arkani-Hamed:2019ymq} also known as the $\sqrt{\textrm{Kerr}}$. The origin name lies in the fact that the Kerr black hole three-point amplitude can be obtained from the double copy of $\sqrt{\textrm{Kerr}}$. Later, it was also shown that for the scattering of black holes \cite{Aoude:2020onz, Vines:2017hyw,Guevara:2018wpp, Arkani-Hamed:2019ymq}, only this UV minimal combination appears. Thus, knowing the Compton amplitude (upto gauge invariant contact terms) for this UV minimal combination would suffice, which has become the reason for a lot of interest in (gravitational-) Compton amplitude in recent times. Even though a great deal of progress has been made, the amplitude suffers from spurious poles whose cancellations require additional contact terms \cite{Cangemi:2023ysz,Arkani-Hamed:2017jhn}. However, it comes with an ambiguity that one can always add local gauge invariant contact terms. There is still no general consensus for what these additional gauge invariant contact terms are. In the effective-one-body (EOB) approach, these contact terms are associated with Wilson coefficients \cite{Levi_2015,Chung:2018kqs,Haddad:2023ylx}.
 
 This brings us to another aspect of Compton scattering which is to describe the scattering of general spinning objects\cite{Bern_2021,Chen:2021kxt,Bern_2023,Kim_2023,Haddad:2023ylx} like neutron stars. There is an abundance of such spinning compact objects that are characterized by spin-induced multipole moments encoding their finite-size structure. Their classical dynamics is captured by the operators in the EOB Hamiltonian that capture these multipole moments. In \cite{Haddad:2023ylx}, the graviton Compton amplitudes for generic three-point interactions have been computed but only up to the sub-leading order of $\hbar$. The authors expanded the generic three-point interactions upto sub-leading order of $\hbar$, expressing it in terms of classical spin and the spin-induced multipole coefficients. These three-point amplitudes were then glued using the BCFW recursion to compute the classical Compton amplitudes. The computations for the full quantum amplitudes (i.e. to arbitrary order of $\hbar$) remain an open problem. In this work, we compute the local, gauge invariant and unitary Compton amplitudes without any restriction on the order of $\hbar$. Although the method we will use is straightforward, it is challenging to write down a closed-form analytical expression of the Compton amplitude for any spin which is local, gauge invariant and unitary.
 
  Another motivation to study Compton scattering for couplings other than the AHH minimal is the UV completion of the higher-spin theories. As a step towards a UV-complete local theory, writing down the perturbative rules of higher-spin EFT involves a deeper understanding of the generic Compton scattering amplitudes. Recently there have been works\cite{Cangemi:2023ysz,Cangemi:2023bpe} where EFT actions for higher-spin particles were written using the Zenoviev formulation of massive gauge symmetry\cite{zinoviev2001massivehighspinparticles,Zinoviev2006OnMS,Zinoviev_2009} and the chiral higher-spin approach\cite{Ochirov_2022}. The authors then used these methods to compute the candidates for higher-spin quantum Compton amplitudes whose classical limit would describe the root-Kerr and Kerr black hole scattering. However, an EFT of higher-spin particles which can describe a generic spinning object with higher multipole moments is yet to be found.
  
  The generic Compton scattering as much as it is relevant for super-heavy and highly spinning compact objects, it has also found its place in probing the hadronic resonances\cite{Pascalutsa_2003,Pascalutsa_2008,L_VOV_2001}. These are the short-lived excitations observed during hadronic collisions in various QCD processes. Experimentally it has been observed that these very short-lived states have spins which lie on Regge trajectories. Since a photon couples directly to the higher-spin charged states, the Compton scattering becomes a very good probe in describing these hadronic excitations. See \cite{Holstein_2014,Wen_2025} for the EFT approaches used in computing the electromagnetic polarizabilities of hadrons from the Compton amplitudes.
  
  Although the gravitational Compton amplitude is relevant for the gravitational-waves scattering conceptually/theoretically, a more tractable problem is first understanding electromagnetic Compton amplitudes. In this paper, we mainly focus on the $2-2$ exchange diagrams for the electromagnetic Compton scattering and the problem of finding the appropriate contact term that restores gauge invariance and unitarity. In this paper, we mainly restrict ourselves to the coupling which is unique in the IR. The motivation behind this is that this coupling is the simplest (contains the least number of derivatives) and has the leading contribution in the IR limit. The existing literature \cite{Cangemi:2023ysz,Arkani-Hamed:2017jhn} mostly focuses on the BCFW/ amplitude factorization approach to compute Compton amplitudes (for the UV minimal coupling) which gives a gauge invariant amplitudes but are not manifestly local and requires additional contact terms to restore locality. In \cite{Kumar:2025juz}, the authors (=three authors from this paper along with Manav Shah) took the more traditional Feynman diagrammatic approach which involves gluing the three-point amplitudes using the higher-spin propagators and is manifestly free of spurious poles. It was shown that the contributions from the exchange diagrams are not gauge invariant, but it is possible to find the suitable contact term to make the total contribution gauge invariant. In that process, the authors also made observations about the analogue of the $R_\xi$ gauge for higher-spin propagators and their relevance in Compton amplitude. The analysis in that paper used variables, which are Lorentz vectors/tensors; there was no reference to the spinor helicity variables.
 
 In this work, we follow the same line of thought. Considering the importance of spinor helicity in the current literature on gravitational waves, it is useful to extend the methods to spinor helicity variables without any reference to Lorentz vectors/tensors. We found that the problem of finding contact terms is much more easily tractable in the spinor helicity variables. The main text has all the computational details. For the benefit of readers, we start by summarizing the result in a self-contained manner. 
 
 \subsection{Main results}
 \label{subsecmainresult}
 Here we summarize the main results of this paper:
 \begin{enumerate}
 	\item Poincar\'e invariance and unitarity uniquely fix the interaction of any charged higher-spin particle in the infrared \cite{Weinberg:1964ew, Weinberg:1964kqu}. In the spinor-helicity (SH, hereafter) formalism, the interaction term is given by 
 	\begin{equation}
 		x \prod\limits_{i=1}^{\spin} \langle 1^{\bfi_i} 2^{\bfj_{j}} \rangle [1^{\bfi_{i+1}} 2^{\bfj_{j+1}}]
 		\label{krssummary1}
 	\end{equation}
 	Here $1$ and $2$ are massive particles with spin $\spin$, and $3$ is the massless particle. The information about particle $3$ is hidden in $x$; this is the $x$ factor introduced in \cite{Arkani-Hamed:2017jhn} \footnote{We do not write down the expression of $x$ here as it is not necessary to communicate the key point of this paper. The explicit expression can be found in \eqref{xfactor1}.}. The soft-photon theorem is governed by this vertex. For future purposes, we refer to this as Weinberg minimal or IR minimal interaction. In the (SH) literature, one more ``minimal" interaction is very popular due to its applicability to Black Hole physics \cite{Arkani-Hamed:2017jhn, Arkani-Hamed:2019ymq,Aoude:2020onz} also known as the AHH minimal. The interaction term in \eqref{krssummary1} is different from AHH minimal\footnote{We thank R Loganayagam for discussions on this point.}. The AHH interaction is the dominant interaction at high energies/short distance; the interaction in \eqref{krssummary1} is unique at low energies/large distance and gives rise to the Coulomb force between two charged particles.

 	\item In the SH formalism, one needs to choose a reference vector to write down the polarizations of a massless particle. This is essentially a choice of a gauge fixing condition. Any physical/observable quantity cannot depend on the gauge-fixing condition, which also means it should not depend on the choice of a reference vector. This is the incarnation of gauge invariance in the SH formalism. Even though this is known in the textbooks \cite{Elvang2013, Schwartz:2014sze}, there is not much literature which uses this principle in the context of scattering amplitude, especially Compton amplitude. In this work, we demonstrate how to use it effectively to fix the contact terms for Compton amplitude. 
 	
 	In the current literature, the opposite helicity Compton amplitude for any spin have been computed for the UV minimal three-point interaction (see eqn 3.14 of \cite{Arkani-Hamed:2017jhn}; we have converted the expression into our notation where $1$\& $2$ are massive and $3$ \& $4$ are massless)
 		\begin{equation}
 			\frac{(\langle {1^{\bfi}}4\rangle[ {2^{\bfj}}3]+\langle {2^{\bfj}}4\rangle[ {1^{\bfi}}3])^{\spin}\langle4|1|3]^{2-2\spin}}{(t+m^2)(u+m^2)}
 		\end{equation}	
 		this expression has poles in $\langle 4|1|3]$ for $\spin>1$ which can be canceled by an appropriate contact term as shown in \cite{Arkani-Hamed:2017jhn}. There are ambiguities regarding what are these contact terms because one can always add an extra gauge invariant contact term. In this work, we propose an algorithm which is manifestly free of spurious poles. We demonstrate them for the Compton amplitude with minimal IR interactions. The IR minimal is the only three-point which is gauge invariant on-shell but not off-shell and thus it needs contact term to ensure gauge invariance. But the same algorithm can be used for any three-point amplitude.

 	\item We compute the tree-level four-point exchange contributions due to this interaction term. It takes the following form 
 	\begin{equation}
 		\widetilde{\mathcal{M}}=	\frac{\mathcal{F}^{(t)}(h_3,h_4)\mathcal{N}^{(t)}(\mathcal{S})	}{t+m^2}
 		+\frac{\mathcal{F}^{(u)}(h_3,h_4)\mathcal{N}^{(u)}(\mathcal{S})	}{u+m^2}
 		\label{krssummary2}
 	\end{equation}
 	where 
 	\begin{eqnarray}
 		\mathcal{F}^{(t)}(h_3,h_4)&=&x_{41}(h_4)\,x_{32}(h_3)
 		\label{krssummary3}
 		\\
 		\mathcal{N}^{(t)}(\mathcal{S})&=&\Omega(\mathcal{S})+\langle 1^{\bfi_1}| 4 |1^{\bfi_{2}}]\langle 2^{\bfj_1}| 3 |2^{\bfj_{2}}]\,  \Upsilon^{(t)}(\mathcal{S})	
 		\label{krssummary4}
 	\end{eqnarray}
 	The explicit expression of $\Omega(\mathcal{S})$ and  $\Upsilon^{(t)}(\mathcal{S})$ are available in the main text as well as in appendix \ref{sec:krsnotationandconv}. The key point is that the subsequent analysis holds true irrespective of the explicit form of $\Omega(\mathcal{S})$ and  $ \Upsilon^{(t)}(\mathcal{S})$, and thus, this method can be generalized to other Compton amplitudes as long as they have the same schematic form. This is one of the key results of the paper. 
 	
 	\item The sum of the exchange diagrams in \eqref{krssummary2} is not independent of the choice of reference vector (and thus not gauge invariant). It is possible to make it gauge invariant after adding a local contact interaction. This local term is not gauge invariant, and the gauge non-invariance of this contact term cancels the gauge non-invariance of the exchange amplitudes in \eqref{krssummary2}. In the main body, we write down the method to find a suitable contact term; the method is very general and can also be applied to gravitational Compton and higher point amplitudes.
 	
 	\item After adding the suitable contact term, the amplitude does not depend on the choice of reference vector. Moreover, using the Schouten identity and a few other identities (momentum conservation, tracelessness \& transversality of the polarizations), it is possible to show that amplitude becomes manifestly independent of the reference vector (i.e. the reference vector drops out from the expressions). As discussed in point 2 above, there are spurious poles for spin $\spin\geq 2$ in the opposite helicity AHH amplitude which needs contact term completion to remove them. In our case, for $h_3=-1$ and $h_4=+1$ helicities of photons, the spurious pole free amplitude takes the following form  
 		\begin{eqnarray}
 			\Bigg[\frac{\Omega(\mathcal{S})}{(t+m^2)(u+m^2)}\frac{\langle 314]^2}{m^2}\,
 			+\frac{\Upsilon^{(t)}(\mathcal{S})}{(t+m^2)}\frac{[1^{\bfi_1}4]^2\langle 2^{\bfj_1}3\rangle^2}{m^4}
 			+\frac{\Upsilon^{(u)}(\mathcal{S})}{(u+m^2)}\frac{[2^{\bfj_1}4]^2\langle 1^{\bfi_1}3\rangle^2}{m^4}
 			\Bigg]
 		\end{eqnarray}
 		where the expression for $\Upsilon^{(t)} $ is given in \eqref{emweinbergminimal15.2}, and $ \Upsilon^{(u)}=\left.\Upsilon^{(t)}\right|_{1\leftrightarrow2}$. We also verify the same with the amplitude factorization method in section \ref{sec:Comparisonwithliterature}.
 	
 	\item Our work is in the context of perturbative effective theory. In general, the higher-spin particles can be a bound state of more fundamental ingredients. Our work is valid as long as the total energy in the c.o.m frame is smaller than the EFT scale\footnote{One example of such a situation could be Large $N$ gauge theory coupled to photons; the large $N$ gauge theory has many flavour singlet higher-spin states.}.

 	\item A natural generalisation of the above scenario is to consider multiple flavours of the massive and massless particles. Let there be $N_F$ number of massive particles and $N$ massless spin 1 particles. We use the following indices
 	\begin{equation}
 		\mathfrak{i}, \mathfrak{j}=1,\cdots, N_F
 		\qquad,\qquad
 		A,B =1, \cdots, N	
 	\end{equation}
 	In this scenario, we can show that the three-point amplitudes involves constants which are of the form $T^A_{\mathfrak{i}\mathfrak{j}}$. From unitarity one could show that these constants satisfy an identity which resembles the form of the structure constants of a Lie algebra\footnote{In \cite{Quevedo:2024kmy} this has been shown for scalar massive particles; here we show it for arbitrary massive spin $\mathcal{S}$ particles.}.
 	
 	\item It is possible to obtain massive spin 1 particles through the Higgs mechanism. When we quantise the theory in the $R_\xi$ gauge, we know that the propagator of the massive spin 1 particle depends on $\xi$. Moreover, the theory also has unphysical ``would-be-goldstone" particles, which ensures unitarity. In \cite{Kumar:2025juz}, the authors considered an ad-hoc extension of that idea to higher-spin particles. In this work, we check the applicability of that idea in the SH formalism. We found that for higher-spin particles, it is possible to find unitary amplitude only for $\xi=1$ and $\xi=\infty$. For $\xi=1$, we need ghosts of spin $0$ to $\mathcal{S}-1$ with three-point amplitudes with specific coefficients.

 	Spinor helicity computations are difficult to do by hand, prone to (minus sign) errors due to wrong application of Schouten identity. Even though there are Mathematica packages for massless SH computations \cite{Maitre:2007jq}, no such package is available for massive SH variables. We developed a package from scratch to handle massive spinor helicity variables, and the computations in this paper are done with the help of our in-house package \cite{kramathematicapscksge}. 
 \end{enumerate}
 \paragraph{Organization of the paper}: In sec \ref{sec:krsbasics}, we review the basics of massive SH formalism. We discuss propagators of spinning massive particles and three-point amplitudes. We also discuss the appearance of reference vectors in the SH variable expressions. In sec \ref{sec:krsfourpoint}, we compute the $t$-channel and $u$-channel exchange diagrams from the Compton amplitude. Then, we check gauge invariance and find the contact term. This section ends with the expression of full gauge-invariant amplitude, which does not depend on any reference vector. In sec \ref{sec:krsgluon}, we extend the results to the case of $N_F$ flavours of massive particles and $N$ flavours of massless spin $1$ particles. In this case, the non-abelian gauge symmetry is important to ensure unitarity. The authors of \cite{Kumar:2025juz} showed that for ensuring unitarity at $\xi=1$ for spin-$\spin$ Compton, one needs to add exchange of particles with spin $\spin-1$ to $0$; the situation is very similar to the exchange of goldstone-like ghosts in Higgsed theory. That analysis has been extended to the SH formalism in sec \ref{sec:krsphotonghost}. In sec \ref{sec:Comparisonwithliterature}, we verify our results with the amplitude factorization method and make a comparison of the IR and UV minimal Compton amplitudes. In sec \ref{sec:krsconclusion}, we discuss the applicability of the method and its usefulness to other cases. We end the section with a few questions that we would like to address in the near future. We have three appendices to assist the readers. In appendix \ref{sec:krsnotationandconv}, we write the notation and the convention for this paper. This also includes the expressions of Mandelstam variables in the SH formalism, the list of symbolic expressions which appear in the main text and few important relations between them. In appendix \ref{krs:conversionformulae}, we provide the conversion rules between Lorentz vectors/tensors and the corresponding quantities in  terms of the SH variables. In appendix \ref{app:offshellterms}, we provide an example for two three-point vertices which differ off-shell but are same on-the shell and gives the same residue in an amplitudes.

 \section{Basics}
 \label{sec:krsbasics}
 We use the massive SH formalism introduced in \cite{Aoude:2020onz, Arkani-Hamed:2017jhn}. It is an extension of the massless SH formalism \cite{Elvang:2015rqa}. We are following the convention given in the book \cite{Elvang:2015rqa}. All the formulae in the paper (especially the minus signs) are sensitive to the convention.   The key variables are 2-dimensional spinors 
 \begin{equation}
 	|\lambda]_\alpha \quad,\quad[\lambda|^\alpha\quad,\quad\langle \lambda|_{\dot \alpha}
 	\qquad,\qquad 
 	|\lambda\rangle^{\dot \alpha}
 	\label{shbasics1}
 \end{equation}
 $\alpha$ and $\dot \alpha$ take two values\footnote{We know that $\textrm{sl}(2\mathbb{C})$ is isomorphic $su(2)_L\times su(2)_R$. Undotted indices ($\alpha,\beta$) are $SU(2)_L$ fundamental indices and dotted indices ($\dot\alpha,\dot \beta$) are $SU(2)_R$ fundamental indices.}. These variables satisfy the Schouten identity
 \begin{eqnarray}
 	\langle r|_{\dot{\alpha}}\langle s|_{\dot{\gamma}}-\langle r|_{\dot{\gamma}}\langle s|_{\dot{\alpha}}&=&-\langle rs\rangle\,\epsilon_{\dot{\alpha}\dot{\gamma}},
 	\qquad,\qquad
 	|r]_{{\alpha}}|s]_{{\gamma}}-|r]_{{\gamma}}|s]_{{\alpha}}=[rs]\,\epsilon_{{\alpha}{\gamma}}	
 	\label{shbasics2}
 \end{eqnarray}
 Here $\epsilon$ is an alternating unit tensor with two indices. 
 Schouten identity plays a central role in any SH computation. The basic idea of SH formalism is that a null vector in $3+1$ dimensions can be written as a product of two spinors
 \begin{eqnarray}
 	p_{\alpha \dot \alpha}=-|p]_{\alpha}\langle p|_{\dot \alpha }
 	\label{shbasics4}
 \end{eqnarray}
 A time-like vector can be written as the sum of two null vectors, and thus, it can be written as\footnote{$p_{\alpha \dot \alpha}$ is a $2\times 2$ matrix. For a null vector, it is a rank 1 matrix and for a time-like vector, it is a rank 2 matrix. We know that a rank-2 matrix can be written as the sum of 2 rank-1 matrices. This is another way to arrive at the SH representation of a time-like momenta.}
 \begin{equation}
 	p_{\alpha\dot \alpha }=-|p^I]_{\alpha}\langle p_I|_{\dot \alpha} 
 	\qquad,\qquad I=1,2
 	\qquad\qquad.
 	\label{shbasics5}
 \end{equation}
 The above expression is invariant if $|p^I]$ transform as a two-dimensional representation of $SU(2)$ and $\langle p_I|$ transform in the conjugate representation. This $SU(2)$ is identified with the little group for massive particles in $3+1$ dimensions. Here, we summarise a few identities which will be very useful in our computation 
 \begin{eqnarray}
 	&&\langle X^{I1}\,X^{I2}\rangle=	 -m_X\epsilon^{I1\,I2}
 	\qquad,\qquad
 	[ X^{I1}\,X^{I2}]= m_X\epsilon^{I1\,I2}
 	\label{shbasics11}
 	\\
 	&&
 	|X_{I1}\rangle^{\dot \alpha}|X^{I1}\rangle ^{\dot \beta} = -m_X\epsilon^{\dot \alpha\dot \beta}
 	\qquad,\qquad
 	|X_{I1}]_{\alpha}|X^{I1}]_{\beta} = m_X\epsilon_{\alpha\beta}
 	\label{shbasics12}
 \end{eqnarray}
 $[a^I|$ is a massive SH variable and $X,Y$ are any two SH variables, then using Schouten identity we can show
 \begin{equation}
 	[a^IX][a_IY]=-m_a[X\,Y]
 	\label{shbasics13}
 \end{equation}	
 We emphasise the fact that every massive particle has its own little group. There is no relation between the little group of two different particles, and thus, little group indices for two different momenta cannot be extracted. For example, consider
 \begin{equation}
 	[X^I\,Y^J]\qquad,\qquad X\ne Y
 	\label{shbasics14}
 \end{equation}
 Even though $I$ and $J$ are $SU(2)$ fundamental representation index, they cannot be contracted since $SU(2)_X$ is different from $SU(2)_Y$. We have used different flavours for indices belonging to different particles to make this visually explicit for the readers. 
 \begin{equation}
 	[X^{\bfi}\,Y^{\bfj}] 
 	\label{shbasics14.1}
 \end{equation}
 Now, we want to briefly discuss the states of massless and massive particles. In $3+1$ dimensions, any massless particle (obeying parity symmetry) has two states; these are denoted by the helicities of the particle. In this paper, we are focusing on massless spin one particles. The two states are denoted by two polarizations $\varepsilon_{i}^{(+1)}$ and $\varepsilon_{i}^{(-1)}$. The expressions of the polarizations in the SH formalism are given by\footnote{We choose the normalization
 	\begin{equation}
 		\varepsilon_{i}^{(+1)}\cdot \varepsilon_{i}^{(-1)}=1	
 		\qquad,\qquad
 		\varepsilon_{i}^{(-1)}\cdot \varepsilon_{i}^{(-1)}=0
 		=\varepsilon_{i}^{(+1)}\cdot \varepsilon_{i}^{(+1)}
 	\end{equation}
 } 
 \begin{equation}
 	[\varepsilon_{i}^{(+1)}(r_i)]_{\alpha\dot{\alpha}}
 	=\sqrt{2}\frac{|i]\langle r_i|}{\langle i\,r_i\rangle }
 	\qquad,\qquad  
 	[\varepsilon_{i}^{(-1)}(r_i)]_{\alpha\dot{\alpha}}
 	=\sqrt{2}\frac{|r_i ]\langle i|}{\langle r_i\,i\rangle }
 	\label{shbasics15}
 \end{equation}
 $r_i$ are called reference vectors. When we try to write down polarizations in the SH formalism, we need to choose one reference vector for each massless leg. These are null vectors and not parallel to the momenta of that leg. For example, if $r_i$ is the reference vector for $i$th leg then 
 \begin{equation}
 	\langle r_i\,i\rangle\ne 0 
 	\qquad,\qquad
 	[r_i\,i]\ne 0
 	\label{shbasics21}
 \end{equation} 
 A key point is that any physical quantity does not depend on the choice of reference vector. Let's try to understand this point better. The SH variables form two-dimensional (complex) vector spaces; hence, at most, two vectors can be linearly independent. From the definition of the reference vector, we know that $r_i$ and $i$ are always linearly independent. Let's say that $r_i^\prime $ is a new reference vector, then 
 \begin{equation}
 	|r_i^\prime\rangle =z_1 |r_i\rangle+z_2 |i\rangle \qquad,\qquad z_1,z_2\in \mathbb{C}\qquad z_1\ne 0 
 	\label{shbasics22}
 \end{equation}
 The difference between polarization vectors corresponding to two reference vectors is essentially a pure gauge polarization, i.e. a gauge transformation (see appendix \ref{sec:krsreferencevecappendix}). Thus, any physical(i.e. gauge invariant) observable should be independent of the choice reference vector. When we compute the Compton amplitude, we do it as a function of the reference vector; then, we demand the four-point amplitude to be independent of the reference vector \footnote{The amplitude is not observable, the cross-section is. So, an amplitude can depend on a gauge through most phases; any other gauge-dependence of amplitude will also make the cross-section gauge dependent. We will see that, in this case, the amplitude changes by an additive term under the change of the reference vector.}. This enables us to find contact terms.

 The states of a massive particle are irreducible representations of little group $SU(2)$; irreducible representations are rank $\mathcal{S}$ tensors with symmetric indices $\{I_{i}\}$ (Each $I$ can take only two values)
 \begin{equation}
 	A^{(I_1\cdots I_{\mathcal{S}})}
 	\label{shbasics23}
 \end{equation}
 these have $\mathcal{S}+1$ independent components. So a massive spin $\mathcal{S}$ particle has $2\mathcal{S}+1$ number of independent states; it is a rank $2\mathcal{S}$ symmetric tensor\footnote{Most textbooks use the language of Lorentz vectors and tensor. In that language, a spin $\mathcal{S}$ particle is a symmetric traceless rank $\mathcal{S}$ tensor.}. For example, polarization of a massive spin one particle (of mass $m$) is given by 
 \begin{equation}
 	\sqrt{2}N\frac{|a^{(\bfi_1}]\langle a^{{\bfi_2)}}|}{m}
 	\qquad,\qquad
 	|a^{(\bfi_1}]\langle a^{{\bfi_2)}}|\equiv 
 	\frac{|a^{\bfi_1}]\langle a^{{\bfi_2}}|+|a^{\bfi_2}]\langle a^{{\bfi_1}}|}{2}
 	\label{shbasics24}
 \end{equation}
 $N$ is the normalization of the polarization vectors. It is given by 
 \begin{equation}
 	N=\frac{1}{2}\delta_{\bfi_1,1}\delta_{\bfi_2,1}+\frac{1}{2}\delta_{\bfi_1,2}\delta_{\bfi_2,2}+\frac{1}{\sqrt{2}}\delta_{\bfi_1,1}\delta_{\bfi_2,2}
 \end{equation}
 $N$ ensures that a massive polarization obeys the following normalization
 \begin{equation}
 	\zeta^{(+)}\cdot 	\zeta^{(-)}=1=\zeta^{(0)}\cdot 	\zeta^{(0)}
 \end{equation}
 Other dot products are zero. In the rest of the paper, we drop $N$ for simplicity. 
 
 \subsection{Propagator of a massive particle}
 \label{subsec:krsbasicsI}
 
 In this section, we discuss the propagator of a massive spinning particle. The propagator in SH formalism is constructed out of the following object 
 \begin{equation}
 	\Theta^{\alpha \dot\alpha \beta \dot\beta}(p) = -2\epsilon^{\alpha\beta} \epsilon^{\dot\alpha \dot\beta}  + \frac{p^{\dot\alpha \alpha} p^{\dot\beta \beta}}{m^2}
 	\label{SHpropagator1}
 \end{equation}
 The propagator is given by \cite{Ingraham:1974un}
 \begin{equation}
 	\frac{-\iimg}{p^2+m^2-\iimg \epsilon}\mathcal{P}_{(\mathcal{S})}^{\{\alpha,\dot \alpha\},\{\beta, \dot \beta \}}
 	\label{SHpropagator2}
 \end{equation}
 $\mathcal{P}_{(\mathcal{S})}^{\{\alpha,\dot \alpha\},\{\beta, \dot \beta \},}$ is the projector for spin $\mathcal{S}$. It is given by 
 \begin{eqnarray}
 	\sum\limits_{a=0}^{\lfloor \frac{\mathcal{S}}{2}\rfloor} \tilde A(\mathcal{S},a)\Theta^{\alpha_1 \dot\alpha_1 \alpha_2 \dot\alpha_2}\Theta^{\beta_1 \dot\beta_1 \beta_2 \dot\beta_2} ... 
 	\Theta^{\alpha_{2a-1} \dot\alpha_{2a-1} \alpha_{2a} \dot\alpha_{2a}}\Theta^{\beta_{2a-1}\dot\beta_{2a-1}\beta_{2a} \dot\beta_{2a}}
 	\Theta^{\alpha_{2a+1} \dot\alpha_{2a+1} \beta_{2a+1} \dot\beta_{2a+1}}...\Theta^{\alpha_{2\mathcal{S}} \dot\alpha_{2\mathcal{S}} \beta_{2\mathcal{S}} \dot\beta_{2J}}	
 	\nonumber\\
 	\label{SHpropagator3}
 \end{eqnarray}
 Here, $\tilde A(\mathcal{S},a)$ is given by 
 \begin{equation}
 	\tilde A(\mathcal{S},a)=\left(-\frac{1}{2}\right)^a\frac{ \mathcal{S}! (2 \mathcal{S}-2 a-1)\text{!!}}{a! (2 \mathcal{S}-1)\text{!!} (\mathcal{S}-2 a)!}
 	\label{SHpropagator11}
 \end{equation}
 These coefficients can be determined by demanding that the propagator is on-shell transverse and traceless. As discussed in \cite{Kumar:2025juz}, one can introduce the $\xi$ parameter.
 \begin{equation}
 	\widetilde{\Theta}^{\alpha \dot\alpha \beta \dot\beta}(p,\xi) = -2\epsilon^{\alpha\beta} \epsilon^{\dot\alpha \dot\beta}  + (\xi-1)\frac{p^{\dot\alpha \alpha} p^{\dot\beta \beta}}{p^2+\xi m^2}
 	\qquad,\qquad \xi \in [0,\infty)	
 	\label{SHpropagator12}
 \end{equation}
 This is motivated by the $\xi$-parameter of $R_\xi$ gauge for Higgesed spin 1 particles. In the case of spin 1, it is possible to derive this expression from gauge-fixing. However, for $\mathcal{S}\geq 2$, we do not have any derivation of this from a Lagrangian. In  \cite{Cangemi:2023ysz}, the authors\footnote{We thank Alexander Ochirov for pointing out this reference.} have given a spin-$\spin$ Lagrangian following the St\"uckelberg mechanism in $D=3+1$ spacetime. After appropriate gauge fixing, one obtains the propagator $\mathcal{P}_{(\spin)}$ which can be derived using the generating function\footnote{Our ansatz at $\xi=1$ (analogous to the Feynman gauge) matches with this form of the propagator for any spin-$\spin$ in $D=2+1$ spacetime.} 
 \begin{align}
 	\mathcal{P}(X,\bar{X})=\sum_{\spin=0}^{\infty}(X)^\spin \cdot \mathcal{P}_{(\spin)} \cdot (\bar{X})^{\spin}=\frac{-\iimg}{p^2+m^2-\iimg \epsilon} \left(\frac{1-\frac{1}{4}\spp{X}{X}\spp{\bar{X}}{\bar{X}}}{1-\spp{X}{\bar{X}}+\frac{1}{4}\spp{X}{X}\spp{\bar{X}}{\bar{X}}}\right)
 \end{align}
 where $X,\bar{X}$ are arbitrary rank $1$ objects. The minus signs have been adjusted as we use mostly positive signature for the metric compared to the mostly negative signature metric in \cite{Cangemi:2023ysz}.

 For $\xi=0$, the propagator is also transverse off-shell. It was shown in \cite{Kumar:2025juz} that the Compton amplitude for spin two and higher-spin particles is gauge invariant only for $\xi=1$ and $\infty$. At $\xi=1$, we need a tower of ghosts (analogue of the ``would-be'' goldstone boson) to make the answer unitary. These might also be analogous to the tower of auxiliary fields which appear in the St\"uckelberg mechanism as shown in \cite{Cangemi:2023ysz,zinoviev2001massivehighspinparticles}. In this work, we restrict to $\xi=\infty$; in that case, the propagator is only on-shell traceless and transverse. In Higgesed theory, this is also known as the unitary gauge propagator. 
 
 \subsection{The $x$ factor}
 \label{subsec:krsbasicsII}
 
 In SH formalism, one variable is important when writing down a three-point amplitude involving a massless particle. It is known as the $x$ factor and is defined as
 \begin{eqnarray}
 	x_{ia}(h_i)=\frac{ \langle r_i|a|i]}{m\langle i\,r_i\rangle }\delta_{h_i,1}-\frac{ \langle i|a|r_i]}{m[ i\,r_i] }\delta_{h_i,-1}
 	\qquad\quad  i\in \textrm{massless leg(s)}
 	\quad,\quad  a\in \textrm{massive leg(s)}
 	\label{xfactor1}	
 \end{eqnarray}
 $r_i$ is the reference vector for the $i$th leg; it is necessary to define the polarization of massless particles in SH formalism. $h_i$ is the helicity of the massless particle. In any gauge-invariant observable, the choice of reference vector does not make any difference. This is the statement of gauge invariance in the SH formalism. We define 
 \begin{eqnarray}
 	\widetilde x_{ia}(h_i)\equiv x_{ia}(h_i)\Big|_{r_i\rightarrow \tilde r_{i}}=\frac{ \langle \tilde r_{i}|a|i]}{m\langle i\,\tilde r_{i}\rangle }\delta_{h_i,1}-\frac{ \langle i|a|\tilde r_{i}]}{m[ i\,\tilde r_{i}] }\delta_{h_i,-1}
 	\label{xfactor2}	
 \end{eqnarray}
 The difference between $x_{ia}(h_i)$ and $\widetilde x_{ia}(h_i)$ (after applying Schouten identity) is given by 
 \begin{eqnarray}
 	\Delta x_{ia}(h_i)\equiv x_{ia}(h_i)-\widetilde x_{ia}(h_i)
 	=
 	\frac{\langle i|a|i]}{m}
 	\Bigg[\frac{\langle r_{i}\,\tilde r_{i}\rangle
 	}{\langle i\,r_{i}\rangle\langle i\,\tilde r_{i}\rangle} \delta_{h_i,1}
 	-\frac{[ r_{i}\,\tilde r_{i}]
 	}{[ i\,r_{i}][ i\,\tilde r_{i}]} \delta_{h_i,-1}
 	\Bigg]
 	\label{xfactor3}	
 \end{eqnarray}
 $i$ is a massless on-shell leg and $a$ is a massive on-shell leg. Thus, the quantity that is sitting outside can be simplified 
 \begin{eqnarray}
 	\langle i|a|i]	=-2k_i\cdot k_a=-\Big( (k_i+k_a)^2+m_a^2\Big) 
 	\label{xfactor4}	
 \end{eqnarray}
 Let's consider a few cases now. We begin with 3-point amplitudes (with $1$ and $2$ being massive and $3$ being massless). In that case $i$ is necessarily $3$, and let's choose $a=1$; then  $(k_i+k_a)^2+m_a^2=m_1^2-m_2^2$. So, it vanishes if the two massive legs have the same mass. Thus, for the three-point amplitude, this is a necessary condition for the $ x$ factor to be gauge-invariant. For four-amplitudes, $(k_i+k_a)^2$ is one of the Mandelstam variables. This will be important to us later.

 \subsection{Three-point amplitudes}
 \label{subsec:krsbasicsIII}
 The three-point interactions in the massive spinor helicity formalism have been extensively discussed in the literature \cite{Conde:2016vxs,Conde:2016izb,Arkani-Hamed:2017jhn} (see also \cite{Basile:2024ydc} in $\text{AdS}_4$ background\footnote{We are thankful to Karapet Mkrtchyan for pointing this out.}).
 For our purpose, we need three-point amplitudes of two massive particles and one massless particle. Furthermore, lets consider that the two massive particles have the same mass. If the spin of the massive leg is $\mathcal{S}$, then there are $2\mathcal{S}+1$ different three-point vertices \cite{Arkani-Hamed:2017jhn}. Among these different vertices, the contribution from one of these vertices dominates in the infrared \cite{Weinberg:1964ew, Weinberg:1964kqu}. It governs the soft photon theorem. We call it IR minimal (or Weinberg minimal). In this work, we focus only on this particular three-point vertex. 
 
 Before writing down the three-point vertex, we introduce $\widehat \otimes_{n}$, which is a symmetrized tensor product. For example, 
 \begin{eqnarray}
 	&&	\Big(\langle 1^{\bfi}|_{\dot \alpha}\Big)^{\widehat \otimes_{n}}
 	=	\langle 1^{(\bfi_1}|_{\dot \alpha_1}\cdots \langle 1^{\bfi_n)}|_{\dot \alpha_n}
 	\label{emweinbergminimal5.1}
 	\\
 	&&	
 	\Big(|1^{\bfi}]_{ \alpha}\langle 1^{\bfi}|_{\dot \alpha}\Big)^{\widehat \otimes_{n}}
 	=	
 	|1^{(\bfi_{1}}]_{ \alpha_1}\cdots
 	|1^{\bfi_{n}}]_{ \alpha_n}
 	\langle 1^{\bfi_{n+1}}|_{\dot \alpha_1}\cdots \langle 1^{\bfi_{2n})}|_{\dot \alpha_n}
 	\label{emweinbergminimal5.2}
 \end{eqnarray}
 The IR minimal three-point amplitude takes the following form 
 \begin{equation}
 	\iimg \tgem \,  \mathcal{C}_{\mathcal{S}}^2\,
 	\frac{(x_{31}-x_{32})}{2}  \Big(\langle 1^{\bfi} 2^{\bfj} \rangle [1^{\bfi} 2^{\bfj}]\Big)^{\widehat \otimes_\mathcal{S} }
 	\qquad,\qquad
 	{\mathcal{C}_\mathcal{S}}^2=\frac{1}{(m^2)^{\mathcal{S}}}	
 	\label{emweinbergminimalthreepoint1}
 \end{equation}
 $x_{31}(h_3) $ is the $x$ factor for particle 3 (see sec. \ref{subsec:krsbasicsII} for discussion on $x$ factor). $\tgem$ is the coupling constant; it has mass dimension 3\footnote{In the Lorentz basis the coupling has the form 
 	\begin{equation}
 		-\iimg \frac{\gem}{2m} (\epsilon_3\cdot k_{12})(\zeta_1\cdot\zeta_2)^{\otimes \mathcal{S}}
 	\end{equation}
 	From the conversion we get $\tgem =(1/\sqrt{2})\,\gem $.
 }. The vertex of a photon with a higher-spin particle picks up a sign under $1\leftrightarrow 2$ exchange; i.e. it is not symmetric in 1 \& 2. In the above equation, we chose a convention that particle 1 has a positive charge and particle 2 has a negative charge. 
 
 For spin 0 and spin 1 particles, in the Lagrangian approach, the above three-point interaction term is generated (along with other interactions) by replacing the derivative with the covariant derivative. As discussed in \cite{zinoviev2001massivehighspinparticles,Cangemi:2022bew}, the similar procedure also generates the minimal coupling for arbitrary massive higher-spin-$\spin$ together with other non-minimal couplings. These works have given a systematic EFT framework for generating local Lagrangians upto quartic order. However, as discussed before a UV-complete and local Lagrangian for higher-spins is yet to be fully developed.

 \subsection{Different $SU(2)$s and their distinctions} In this work, we are focusing on 4-point Compton amplitudes. In our convention, the massive particles are labeled with indices $1$ and $2$. In this problem, there are four different $SU(2)$s. 
 \begin{itemize}
 	\item $SU(2)_L$ comes from Lorentz group. We use undotted indices for this. 
 	\item $SU(2)_R$ comes Lorentz group. We use dotted indices for this. 
 	\item $SU(2)_{(1)}$ is the little group for massive particles labeled with 1. We use $\bfi_i$ as indices. 
 	\item $SU(2)_{(2)}$ is the little group for massive particles labeled with 2. We use $\bfj_i$ as indices. 
 \end{itemize}

 \section{Four-point exchange contributions and contact terms}
 \label{sec:krsfourpoint}
 
 In this section, we compute the Compton amplitude for the minimal coupling that governs the electromagnetic soft theorem (see Fig. \ref{fig:higherspincomptondiag}; there is no $s$-channel diagram as it is not possible to have self-interaction of photons, which also obey Bose statistics). 
 \begin{figure}[h]
 	\begin{center}
 		\begin{tikzpicture}[line width=1.5 pt, scale=1.3]
 			
 			\begin{scope}[shift={(0,0)}]	
 				
 				\treefourpointexchange{chspin}{photon}{photon}{chspinbar}{chspin}{0}	
 				
 				\labeltreefourpointexchange{ $k_2$ }{ $k_3$ }{ $k_4$ }{ $k_1$ }{}{0} 
 				
 				\draw (0,-2.5) node {$t$-channel};
 			\end{scope}

 			\begin{scope}[shift={(7,0)}]	
 				
 				\treefourpointcrossexchange{photon}{chspinbar}{chspin}{photon}{chspin}{180}	
 				
 				\labeltreefourpointexchange{ $k_2$ }{ $k_3$ }{ $k_4$ }{ $k_1$ }{}{0} 
 				\draw (0,-2.5) node {$u$-channel};
 				
 			\end{scope}

 		\end{tikzpicture} 
 	\end{center}
 	\caption{Higher-spin Compton scattering}
 	\label{fig:higherspincomptondiag}
 \end{figure}
 We call it the IR minimal coupling or Weinberg Minimal coupling. We spell out our convention clearly here
 \begin{itemize}
 	\item Particle $1$ \& $2$ are massive; the charge for particle 1 is positive, and the charge for particle 2 is negative. This is also clear from the arrow in the Feynman diagram. 
 	\item Particle $3$ \& $4$ are massless particles.
 \end{itemize}
 The three-point amplitude is given by \eqref{emweinbergminimalthreepoint1}. We start with two such vertices. We strip one of the massive legs from each vertex. For example, in the case of $t$-channel, the stripped vertices\footnote{Stripping one massive leg removes a factor of $\mathcal{C}_{\mathcal{S}}$. } are given by\footnote{From the momentum conservation at the vertices, we can convert the two $x$ factors in \eqref{emweinbergminimalthreepoint1} to a single $x$ factor involving only the external massive leg. One can always use this as long as the massless leg is on-shell. We write the vertices in this section after using this logic. }  
 \begin{equation}
 	-
 	\iimg \tgem\frac{\mathcal{C}_{\mathcal{S}}}{2^{\mathcal{S}/2}}\,
 	x_{32}  \Big(|2^{\bfj_{i}}]_{\alpha}\langle 2^{\bfj_{i+1}}|_{\dot \alpha}\,\Big)^{\widehat \otimes_\mathcal{S} }
 	\qquad,\qquad
 	\iimg \tgem\frac{\mathcal{C}_{\mathcal{S}}}{2^{\mathcal{S}/2}}\,\,
 	x_{41}  \Big(|1^{\bfi_{i}}]_{\alpha}\langle 1^{\bfi_{i+1}}|_{\dot \alpha}\,\Big)^{\widehat \otimes_\mathcal{S} } \label{StrippedVertices}
 \end{equation}
 and glue two such stripped 3-point amplitudes to get the four-point exchange. {In general, the stripped vertices can have more terms which vanish on-shell i.e. when the vertex is contracted back with the stripped legs. We do not include such terms because these extra terms leads to contact terms and so do not contribute at the poles. We discuss a simple example of such terms in the appendix \ref{app:offshellterms}.
 }
 
 In order to compute the amplitude, we compute the scalars that are constructed from the SH variables of the external legs and the projector. Before going into the details of the computation, we begin with the definition of $\odot$ product
 \begin{equation}
 	\big(X\odot_t Y\big)\equiv 	\langle X|^{(\bfi_1}_{\dot\alpha_1} |X]^{\bfi_2)}_{\alpha_1 } 
 	\Theta^{\alpha_1 \dot\alpha_1 \alpha_2  \dot\alpha_2}(k_1+k_4)	
 	\langle Y|^{(\bfj_1}_{\dot\alpha_2} |Y]^{\bfj_2)}_{\alpha_2} 
 	\qquad,\qquad 
 	\big(X\odot_t Y\big)=\big(Y\odot_t X\big)
 	\label{emweinbergminimal2}
 \end{equation}
 We can evaluate it explicitly; it is given by 
 \begin{equation}
 	\big(X\odot_t Y\big)=
 	2\langle X^{\bfi_1}Y^{\bfj_1}\rangle [X^{\bfi_2}Y^{\bfj_2}]
 	+\frac{1}{m^2}
 	\langle X^{\bfi_1}| (1+4) |X^{\bfi_2}]
 	\langle Y^{\bfj_1}| (1+4) |Y^{\bfj_2}]
 	\label{emweinbergminimal3}
 \end{equation}
 Here $\bfi$ indices are all symmetrized, and $\bfj$ indices are all symmetrized. We will not write this explicitly. It will be assumed from now on unless it is stated otherwise. Then the various $\odot_t $ product are given by 
 \begin{eqnarray}
 	&&\big(1\odot_t 1\big)
 	=
 	\frac{1}{m^2}\langle 1^{\bfi}| 4 |1^{\bfi}]^{\widehat \otimes_ 2}	  
 	\label{emweinbergminimal4.1}
 	\\
 	&&
 	\big(2\odot_t 2\big)
 	=
 	\frac{1}{m^2}\langle 2^{\bfj}| 3 |2^{\bfj}]^{\widehat \otimes_2}	
 	\label{emweinbergminimal4.2}
 	\\
 	&&
 	\big(1\odot_t 2\big)
 	=
 	2\langle 1^{\bfi_1}2^{\bfj_1}\rangle [1^{\bfi_2}2^{\bfj_2}]
 	-
 	\frac{1}{m^2}\langle 1^{\bfi_1}| 4 |1^{\bfi_2}]\langle 2^{\bfj_1}| 3 |2^{\bfj_2}] 
 	\label{emweinbergminimal4.3}
 \end{eqnarray}
 Then, the numerator of the $t$-channel is given by 
 \begin{equation}
 	\mathcal{F}^{(t)}(h_3,h_4)\mathcal{N}^{(t)}(\mathcal{S})	
 	\qquad,\qquad
 	\mathcal{F}^{(t)}(h_3,h_4)=x_{41}(h_4)\,x_{32}(h_3)
 	\label{emweinbergminimal11}
 \end{equation}
 $x$-factors are defined in \eqref{xfactor1}. From the definition, it is clear that the helicity dependence is completely captured by $\mathcal{F}^{(t)}(h_3,h_4)$; $\mathcal{N}^{(t)}(\mathcal{S})	$ does not depend on the helicity of the massless particle. Later, we will see the usefulness of writing down the expressions in such a product form. $\mathcal{N}^{(t)}(\mathcal{S})$ is defined as 
 \begin{equation}
 	\mathcal{N}^{(t)}(\mathcal{S})=	\frac{\mathcal{C}_{\mathcal{S}}^2}{2^\mathcal{S} } \Big(|1^{\bfi}]_{ \alpha}\langle 1^{\bfi}|_{\dot \alpha}\Big)^{\widehat \otimes_{\mathcal{S}}}
 	\mathcal{P}_{\mathcal{S}}^{\{\alpha\}\{\dot \alpha\}\{\beta\}\{\dot \beta\}} 
 	\Big(|2^{\bfj}]_{ \beta}\langle 2^{\bfj}|_{\dot \beta}\Big)^{\widehat \otimes_{\mathcal{S}}}
 	\label{emweinbergminimal12}
 \end{equation}
 Here $\mathcal{P}_{\mathcal{S}}$ is the projector that appears in the Massive higher-spin propagator (see \eqref{SHpropagator3}). Then, the total amplitude $\iimg \widetilde{\mathcal{M}}$ is the sum of $t$-channel and $u$-channel exchange. $\widetilde{\mathcal{M}}$ is given by 
 \begin{equation}
 	\widetilde{\mathcal{M}}=	(\tgem)^2\Big(\frac{\mathcal{F}^{(t)}(h_3,h_4)\mathcal{N}^{(t)}(\mathcal{S})	}{t+m^2}
 	+\frac{\mathcal{F}^{(u)}(h_3,h_4)\mathcal{N}^{(u)}(\mathcal{S})	}{u+m^2}\Big)
 	\label{emweinbergminimal12.1}
 \end{equation}
 We begin our analysis with $\mathcal{N}^{(t)}(\mathcal{S})$. From the definition, it is clear that $\mathcal{N}^{(t)}(\mathcal{S})$ does not depend on the helicity of the photon. Putting the expressions of $\mathcal{P}$ from \eqref{SHpropagator3}, we get the following expression of  $\mathcal{N}^{(t)}(\mathcal{S})$
 \begin{equation}
 	\mathcal{N}^{(t)}(\mathcal{S})=	\frac{\mathcal{C}_{\mathcal{S}}^2}{2^\mathcal{S} }\sum_{a=0}^{\lfloor\frac{\mathcal{S}}{2} \rfloor} \tilde A(\mathcal{S},a)\big(1\odot_t 1\big)^{\widehat \otimes_a}\big(2\odot_t 2\big)^{\widehat \otimes_a}\big(1\odot_t 2\big)^{\widehat \otimes_{(\mathcal{S}-2a)}}
 	\label{emweinbergminimal13}
 \end{equation}
 In the above equation, all the little groups over leg $1$ are symmetrized, and all the little groups over leg 2 are symmetrized.
 We will first compute $\mathcal{N}^{(t)}(\mathcal{S})$ explicitly. If we put all the expressions from \eqref{emweinbergminimal4.1}, \eqref{emweinbergminimal4.2} and \eqref{emweinbergminimal4.3}, we get \footnote{In order to arrive at this expression, we first binomially expand $\big(1\odot_t 2\big)^{\widehat \otimes_{(\mathcal{S}-2a)}}$ and then interchange the two summations. This process is explained in detail in \cite{Kumar:2025juz}.}
 \begin{equation}
 	\mathcal{N}^{(t)}(\mathcal{S})=	\mathcal{C}^2_\spin	\sum_{b=0}^{\mathcal{S}}\frac{(-1)^{\mathcal{S}-b}}{m^{2(\mathcal{S}-b)}2^{\mathcal{S}-b}}\tilde B(\mathcal{S},b)
 	\Big(\langle 1^{\bfi_i}| 4 |1^{\bfi_{i+1}}]\langle 2^{\bfj_i}| 3 |2^{\bfj_{i+1}}] \Big)^{\widehat \otimes_{\mathcal{S}-b}}\Big(
 	\langle 1^{\bfi_j}2^{\bfj_j}\rangle [1^{\bfi_{j+1}}2^{\bfj_{j+1}}]
 	\Big)^{\widehat \otimes_b} 
 	\label{emweinbergminimal14}
 \end{equation}
 Here $\tilde B(\mathcal{S},b)$ is given by 
 \begin{equation}
 	\label{comptonminimalcalculation24}
 	\tilde B(\mathcal{S},b) = \sum\limits_{a=0}^{\lfloor \frac{\mathcal{S}-b}{2} \rfloor}  {}^{\mathcal{S}-2a}\mathrm{C}_b \ \tilde A(\mathcal{S},a)
 \end{equation}
{ where ${}^{\mathcal{S}-2a}\mathrm{C}_b $ is the binomial factor.} We rewrite the numerator $\mathcal{N}^{(t)}(\mathcal{S})$ in the following way (it is understood that the little group indices for leg $1$ and $2$ are symmetrized)
 \begin{equation}
 	\mathcal{N}^{(t)}(\mathcal{S})=\Omega(\mathcal{S})+\frac{\langle 1^{\bfi_1}| 4 |1^{\bfi_{2}}]\langle 2^{\bfj_1}| 3 |2^{\bfj_{2}}]}{m^4} \Upsilon^{(t)}(\mathcal{S})
 	\label{emweinbergminimal21}
 \end{equation}
 here we have defined the following variables  
 \begin{eqnarray}
 	\Omega(\mathcal{S})&\equiv&\mathcal{C}_{\mathcal{S}}^2\Big(
 	\langle 1^{\bfi_j}2^{\bfj_j}\rangle [1^{\bfi_{j+1}}2^{\bfj_{j+1}}]
 	\Big)^{\widehat \otimes_\mathcal{S}} 
 	\label{emweinbergminimal15.1}
 	\\
 	\Upsilon^{(t)}(\mathcal{S})&\equiv&	\mathcal{C}_{\mathcal{S}}^2\sum_{b=0}^{\mathcal{S}-1}\frac{(-1)^{\mathcal{S}-b}}{m^{2(\mathcal{S}-b-2)}2^{\mathcal{S}-b}}\tilde B(\mathcal{S},b)
 	\Big(\langle 1^{\bfi_i}| 4 |1^{\bfi_{i+1}}]\langle 2^{\bfj_i}| 3 |2^{\bfj_{i+1}}] \Big)^{\widehat \otimes_{\mathcal{S}-b-1}}\Big(
 	\langle 1^{\bfi_j}2^{\bfj_j}\rangle [1^{\bfi_{j+1}}2^{\bfj_{j+1}}]
 	\Big)^{\widehat \otimes_b} 
 	\label{emweinbergminimal15.2}
 \end{eqnarray}
 In \eqref{emweinbergminimal21}, we have left out a factor of $\langle 1^{\bfi_1}| 4 |1^{\bfi_{2}}]\langle 2^{\bfj_1}| 3 |2^{\bfj_{2}}]$; we will see in the next section that this factor is very useful to ensure gauge invariance. Note that the definition of $\Omega(\mathcal{S})$ is not specific to a particular channel (i.e. it is invariant under $1\leftrightarrow 2$ exchange); $\Upsilon^{(t)}(\mathcal{S})$ and $\Upsilon^{(u)}(\mathcal{S})$  does. $\Upsilon^{(u)}(\mathcal{S})$ and $\Upsilon^{(t)}(\mathcal{S})$ are related by $1\leftrightarrow 2$ exchange.  We have defined $\Omega(\mathcal{S})$ and $\Upsilon^{(t)}(\mathcal{S})$ to be dimensionless.   In all the above expressions, the little group indices over leg 1 and leg 2 are symmetrized.

 Now we want to understand $\mathcal{F}$, defined in \eqref{emweinbergminimal11}. Again, it is evident from the definition that it does not depend on the spin of the massive leg. Secondly, it is not invariant under $1\leftrightarrow 2$ exchange, so it depends on the specific channel. We use the definition in \eqref{xfactor1} to write,
 \begin{align}
 	\mathcal{F}^{(t)}(h_3,h_4) = 
 	\frac{\langle r_{3}|2|3]\langle r_{4}|1|4]}{m^2\langle 3\,r_{3}\rangle \langle 4\,r_{4}\rangle }&\mathfrak{H}_{h_3,h_4}
 	-\frac{\langle r_{3}|2|3]\langle 4|1|r_{4}]}{m^2\langle 3\,r_{3}\rangle [4\,r_{4}] }\mathfrak{H}_{h_3,-h_4}\nonumber\\
 	&-\frac{\langle 3|2|r_{3}]\langle r_{4}|1|4]}{m^2[3\,r_{3}]\langle 4\,r_{4}\rangle }\mathfrak{H}_{-h_3,h_4}
 	+
 	\frac{\langle 3|2|r_{3}]\langle 4|1|r_{4}]}{m^2[3\,r_{3}] [4\,r_{4}] }
 	\mathfrak{H}_{-h_3,-h_4}
 	%
 	\nonumber\\
 	\label{emweinbergminimal22}
 \end{align}
 where $\mathfrak{H}(h_3,h_4)$ is defined in term product of two Heaviside theta function
 \begin{eqnarray}
 	\mathfrak{H}_{h_3,h_4}=H(h_3)H(h_4)
 	\label{emweinbergminimal23}
 \end{eqnarray}
 \subsection{Checking gauge invariance}
 \label{subsec:krsfourpointI}
 
 Now, we want to check the gauge invariance of the full amplitude. In order to do that, we replace the reference vector for leg 3 (or for leg 4) with a new reference vector and compute the difference. We have mentioned that the information of the massless legs is entirely in the $\mathcal{F}$s. Let's check what happens to $\mathcal{F}$ if we change one of the reference vectors
 \begin{eqnarray}
 	\Delta_i\mathcal{F}=\mathcal{F}-\mathcal{F}\Big|_{r_{i}\rightarrow \tilde r_{i}}
 	\qquad,\qquad i\in 3,4
 	\label{emweinbergminimal51}
 \end{eqnarray} 
 Given the fact that $\mathcal{F}$ is a product of two $x$ factors (see \eqref{emweinbergminimal11}), we can use \eqref{xfactor3} to compute the above quantity. For example
 \begin{eqnarray}
 	\Delta_3\mathcal{F}^{(t)}(h_3,h_4)=x_{41}(h_4)\,\Big(\Delta x_{32}(h_3)\Big)=
 	x_{41}(h_4)
 	\frac{\langle 3|2|3]}{m^2}	A_3 \quad\mathrm{and}\quad \Delta_4\mathcal{F}^{(t)}(h_3,h_4) = x_{31}(h_3)\frac{\langle 4|1|4]}{m^2}	A_4
 	\nonumber\\	
 	\label{emweinbergminimal52}
 \end{eqnarray}
 where
 \begin{equation}
 	\label{eq:Adef}
 	A_i \equiv m \Bigg[\frac{\langle r_{i}\,\tilde r_{i}\rangle
 	}{\langle i\,r_{i}\rangle\langle i\,\tilde r_{i}\rangle} \delta_{h_i,1}
 	-\frac{[ r_{i}\,\tilde r_{i}]
 	}{[ i\,r_{i}][ 3\,\tilde r_{i}]} \delta_{h_i,-1}
 	\Bigg]
 \end{equation}
 In this equation we $\langle 3|2|3]=\langle 4|1|4]=(t+m^2)$ to simplify it even further. With this information, let's now check what happens to the amplitude (defined in \eqref{emweinbergminimal12.1}) when we change one of the reference vectors
 \begin{eqnarray}
 	\Delta_i \widetilde{\mathcal{M}}\equiv \widetilde{\mathcal{M}}-\widetilde{\mathcal{M}}\Big|_{r_{i}\rightarrow \tilde r_{i}}	
 	=
 	\Big(\Delta_i\mathcal{F}^{(t)}(h_3,h_4)\Big)\frac{\mathcal{N}^{(t)}(\mathcal{S})	}{t+m^2}
 	+\Big(\Delta_i\mathcal{F}^{(u)}(h_3,h_4)\Big)\frac{\mathcal{N}^{(u)}(\mathcal{S})	}{u+m^2}
 	\label{emweinbergminimal53}
 \end{eqnarray}
 Now we use \eqref{emweinbergminimal52}, along with $\langle 3|2|3]=(t+m^2)$, $\langle 3|1|3]=(u+m^2)$, so we see that poles cancel in the difference. Nevertheless, the difference is non-vanishing, and thus $\widetilde{\mathcal{M}}$, which is the sum of $t$-channel and $u$-channel exchange, is not gauge invariant. Since there is no pole in the difference, the gauge non-invariance of the amplitude can only be cured by a contact term. In order to be certain about that possibility, we need to check whether there is any gauge non-invariant contact term, such that the non-invariance of the contact term cancels the difference above.

 \subsection{Can other exchanges cure the problem ?}\label{sec:CanotherExchanges?}
 
 We found that the sum of exchange diagrams is not gauge invariant. So, a very natural question to ask is whether a massless/massive exchange can cure the problem\footnote{We are thankful to Arnab Priya Saha for this question and for the discussion on this point.}? 
 \begin{itemize}
 	\item Let's first consider the possibility of $t$ and $u$ channel massless exchanges. There exists a non-zero vertex between a massive higher-spin particle and two massless photons \cite{Chakraborty:2020rxf}. However, the exchange diagrams due to that vertex are gauge invariant by itself \cite{Balasubramanian:2021act}; it cannot cure this problem. 
 	\item Let's consider the massless exchange in the $s$ channel. That would need a three-massless vertex. The only non-zero three-massless vertex with two photons is photon-photon-graviton. That cannot resolve this problem \cite{Weinberg:1964ew}. 
 	\item The same argument can be extended to massive exchanges where the mass of the exchange particle is different from the external particles. In that case, the only allowed three-point amplitude is in terms of linearised field strength $\mathcal{B}_{\mu\nu}$ and thus, the contributions from the exchange diagram are again gauge-invariant by themselves.
 	\item This leaves with us the possibility of exchange of particles with the same mass and same/different spin. If the exchange particle has the same mass but a different spin, then the vertex will have a different number of derivatives. {For example : Consider the $2-2$ Compton scattering of a spin $1$ particle with the exchange of a spin $0$ particle. The minimal coupling for spin $1$-spin $0$-photon is given by $\frac{\iimg}{2m^2} (\zeta_1\cdot k_3)(\varepsilon_3\cdot k_{12})$ which has one extra derivative as compared to the spin $1$-spin $1$-photon coupling $\frac{\iimg}{2m} (\zeta_1\cdot \zeta_2)(\varepsilon_3\cdot k_{12})$. Therefore, the numerators of the exchange of spin $0$ particles would have extra momentum factors. For example : the tensor factors in $t$ channel numerator would be $\frac{1}{m^4}(\varepsilon_3\cdot k_2)(\varepsilon_4\cdot k_1)(\zeta_1\cdot k_4)(\zeta_2\cdot k_3)$ which in spinor helicity formalism is given by $x_{32}(h_3)x_{41}(h_4) \frac{\SHAB{1^{\bfi_1}}{4}{1^{\bfi_2}} \SHAB{2^{\bfj_1}}{3}{2^{\bfj_2}}}{m^4}$. The gauge transformation of this term would cancel with the gauge transformation of the second term in the spin $1$ exchange $\mathcal{F}^{(t)}(h_3,h_4)\mathcal{N}^{(t)}(1)$, where $\mathcal{N}^{(t)}(\spin)$ is given in eq. \eqref{emweinbergminimal21}. The first term proportional to $\Omega(\spin)$ would still survive and require additional contact term. This implies that a exchanges with different spin cannot cure the non-gauge invariance without adding a contact term.}
 	
 	Then, the only possibility is that the exchange particle has the same mass and the same spin. This means that we have more than one massive spin $\spin$ particle with the same mass. We discussed the situation in detail in sec \ref{sec:krsgluon}. We will see that even in that case, the problem of gauge non-invariance remains if we do not add a contact term. 
 \end{itemize}
 
 \subsection{Search for  Contact term to restore gauge invariance}
 \label{subsec:krsfourpointII}
 
 In \eqref{emweinbergminimal53}, we can look for the coefficient of  $\Omega(\mathcal{S})$, $\Upsilon^{(t)}(\mathcal{S})$ and $\Upsilon^{(u)}(\mathcal{S})$; it happens to be that one can find a suitable contact term to cancel those contributions and it is possible to do that for each of those terms separately \cite{Kumar:2025juz} \footnote{In terms of the basis, $ \Omega(\mathcal{S}) $, $\Upsilon^{(t)}(\mathcal{S})$ and $\Upsilon^{(u)}(\mathcal{S})$ are independent basis elements.}. Let's suppose that the contact term is of the form 
 \begin{equation}
 	\iimg \widetilde{\mathcal{M}}^{(\textrm{ct})}=	\iimg \frac{\tgem^2}{m^2} \Bigg(  \Omega(\mathcal{S}) \mathcal{G}_{\Omega}
 	+\Upsilon^{(t)}(\mathcal{S})\mathcal{G}_{\Upsilon^{(t)}}+\Upsilon^{(u)}(\mathcal{S})\mathcal{G}_{\Upsilon^{(u)}}\Bigg)
 	\label{emweinbergminimal52.1}
 \end{equation}
 such that 
 \begin{equation}
 	\mathcal{M}\equiv \widetilde{\mathcal{M}}+\widetilde{\mathcal{M}}^{(\textrm{ct})}
 	\label{emweinbergminimal52.2}
 \end{equation}
 is gauge invariant. The goal of this section is to check whether $\widetilde{\mathcal{M}}^{(\textrm{ct})}$ exists or not, and if yes, then what it is.

 First, we explain one general strategy to find the contact term. Say the contact term is $\mathcal{G}(r_3,r_4)$. Then $\Delta_3\mathcal{G}(r_3,r_4)$ is given by 
 $\mathcal{G}(r_3,r_4)-\mathcal{G}(\tilde r_3,r_4)$. Thus in order to find the contact term $\mathcal{G}(r_3,r_4)$ we need to break $\Delta_3\widetilde{\mathcal{M}}$ in \eqref{emweinbergminimal53} into two terms, one involving only $r_3$ and another involving only $\tilde r_{3}$ (and similar conditions for leg 4). From the expression, we can see that $\Delta_3 \widetilde{\mathcal{M}}$ always has one the following two terms 
 \begin{eqnarray}
 	\frac{\langle r_{3}\,\tilde r_{3}\rangle
 	}{\langle 3\,r_{3}\rangle\langle 3\,\tilde r_{3}\rangle}
 	\qquad,\qquad
 	\frac{[ r_{3}\,\tilde r_{3}]
 	}{[3\,r_{3}][ 3\,\tilde r_{3}]}	
 	\label{emweinbergminimal53.1}
 \end{eqnarray}	
 Let's focus on the first term (which is for $h_3=1$). If we can find a term $\langle 3X\rangle$ in the numerator, then we use the following Schouten identity to separate $r_3$ and $\tilde r_{3}$
 \begin{eqnarray}
 	\langle r_{3}\,\tilde r_{3}\rangle\langle 3X\rangle &	=& \langle 3\,r_3\rangle\langle X\,\tilde r_{3}\rangle-\langle 3\,\tilde r_{3}\rangle\langle X\,r_3\rangle
 	\label{emweinbergminimal53.3}
 \end{eqnarray}	
 after this we set all the terms containing $\langle \tilde r_{3},X\ne 3\rangle$ to zero to get the desire contact term. For $h_3=-1$, we replace the angle variables with box variables. For $i=4$, we follow the same strategy with $3$ replaced by $4$.
 
 \subsubsection{Finding $\mathcal{G}_{\Omega}$}
 \label{sec:findingGomega}
 We start implementing the logic in one example. First, we focus on the term proportional to $\Omega(\mathcal{S})$ in \eqref{emweinbergminimal53}. For example, for $i=3$, they are given by 
 \begin{equation}
 	-\frac{1}{m}
 	\Bigg[\frac{\langle r_{3}\,\tilde r_{3}\rangle
 	}{\langle 3\,r_{3}\rangle\langle 3\,\tilde r_{3}\rangle} \delta_{h_3,1}
 	-\frac{[ r_{3}\,\tilde r_{3}]
 	}{[ 3\,r_{3}][ 3\,\tilde r_{3}]} \delta_{h_3,-1}
 	\Bigg](		x_{41}(h_4)+x_{42}(h_4))
 	\label{emweinbergminimal54}
 \end{equation}
 using the expression in \eqref{xfactor1}, the term in parenthesis can be simplified 
 \begin{equation}
 	(x_{41}(h_4)+x_{42}(h_4))= -\frac{1}{m}\Bigg(\frac{\langle r_4|3|4]}{\langle 4\,r_4\rangle }\delta_{h_4,1}-\frac{\langle 4|3|r_4]}{[ 4\,r_4] }\delta_{h_4,-1} \Bigg)
 	\label{emweinbergminimal55}
 \end{equation}
 This falls into the form of LHS of \eqref{emweinbergminimal53.3}, and hence, we can apply that identity. The contact term $\mathcal{G}_{\Omega}(h_3,h_4)$ (defined in \eqref{emweinbergminimal52.1}) is of the following form
 \begin{equation}
 	\mathcal{G}_{\Omega}=\frac{[34]\langle r_{3}\,r_{4}\rangle}{\langle 3\,r_{3}\rangle\langle 4\,r_{4}\rangle}
 	\mathfrak{H}_{h_3,h_4}
 	+
 	\frac{\langle 4\,r_{3}\rangle[3\,r_{4}]}{\langle 3\,r_{3}\rangle[4\,r_{4}]}
 	\mathfrak{H}_{h_3,-h_4}
 	+
 	\frac{\langle 3\,r_{4}\rangle[ 4\,r_{3}]}{[3\,r_{3}]\langle4\,r_{4} \rangle}
 	\mathfrak{H}_{-h_3,h_4}
 	+
 	\frac{\langle 3\,4\rangle[r_{3}\,r_{4}]}{[3\,r_{3}][4\,r_{4}]}
 	\mathfrak{H}_{-h_3,-h_4}
 	\label{emweinbergminimal63}
 \end{equation}
 We can see that if we put $r_{3}=4$ and $r_{4}=3$, then this term vanishes for opposite helicity. So, the contact term makes a difference for amplitudes with the same helicity of the photons. Note that $\mathcal{G}_{\Omega}$ does not depend on the spin of the massive legs. 
 
 After adding the contact terms, the total $\Omega$-dependent amplitude becomes 
 \begin{equation}
 	\mathcal{M}_\Omega \equiv \tgem^2\Omega(\mathcal{S})\Bigg[\frac{\mathcal{F}^{(t)}(h_3,h_4)}{(t+m^2)}+\frac{\mathcal{F}^{(u)}(h_3,h_4)}{(u+m^2)}+\frac{1}{m^2}\mathcal{G}_{\Omega}(h_3,h_4)\Bigg]
 	\label{emweinbergminimal64}
 \end{equation}
 Now we find $\mathcal{G}_{\Upsilon^{(t)}}$ and $\mathcal{G}_{\Upsilon^{(u)}}$ (see \eqref{emweinbergminimal52.1} for definition). The expressions of $\Upsilon$s depend only on one channel, and thus, contributions from two channels do not add up. 
 
 \subsubsection{Finding $\mathcal{G}_{\Upsilon}$}
 
 Let's focus on $\Upsilon^{(t)}$ (terms corresponding to $\Upsilon^{(u)}$ can be obtained by $1\leftrightarrow 2$ exchange). For $\Omega(\mathcal{S})$, the contribution from both the channels was needed to implement \eqref{emweinbergminimal53.3}. Unlike that case, it is simpler to implement the strategy given in \eqref{emweinbergminimal53.3} for $\Upsilon^{(t)}$ and $\Upsilon^{(u)}$ as we already have necessary terms ($\langle 3X\rangle$, $[ 3X]$, $\langle 4, Y\rangle$ \& $[ 4, Y]$) in the numerator for each channel separately; this is why we left-out $\langle 1^{\bfi_1}| 4 |1^{\bfi_{2}}]\langle 2^{\bfj_1}| 3 |2^{\bfj_{2}}]$ in \eqref{emweinbergminimal21}. The necessary contact term  $\mathcal{G}_{\Upsilon^{(t)}}$ is of the following form
 \begin{eqnarray}
 	\mathcal{G}_{\Upsilon^{(t)}}=&&-\Big[\frac{\SHAA{1^{\bfi_1}}{r_4}\SHAB{r_3}{2}{3} \SHAB{2^{\bfj_1}}{3}{2^{\bfj_2}}\SHBB{1^{\bfi_2}}{4}}{m^4\SHAA{3}{r_3}\SHAA{4}{r_4}}+
 	\frac{\SHAA{2^{\bfj_1}}{r_3}\SHBB{1^{\bfi_1}}{4}\SHBB{1^{\bfi_2}}{4}\SHBB{2^{\bfj_2}}{3}}{m^3\SHAA{3}{r_3}}
 	\Big]	\mathfrak{H}_{h_3,h_4}
 	\nonumber
 	\\
 	+&&\Big[\frac{\SHBB{1^{\bfi_1}}{r_4}\SHAB{r_3}{2}{3} \SHAB{2^{\bfj_1}}{3}{2^{\bfj_2}}\SHAA{1^{\bfi_2}}{4}}{m^4\SHAA{3}{r_3}\SHBB{4}{r_4}}+
 	\frac{\SHAA{2^{\bfj_1}}{r_3}\SHAA{1^{\bfi_1}}{4}\SHAA{1^{\bfi_2}}{4}\SHBB{2^{\bfj_2}}{3}}{m^3\SHAA{3}{r_3}}
 	\Big]		\mathfrak{H}_{h_3,-h_4}
 	\label{emweinbergminimal71}
 	\\
 	+&&\Big[\frac{\SHAA{1^{\bfi_1}}{r_4}\SHAB{3}{2}{r_3} \SHAB{2^{\bfj_1}}{3}{2^{\bfj_2}}\SHBB{1^{\bfi_2}}{4}}{m^4\SHBB{3}{r_3}\SHAA{4}{r_4}}+
 	\frac{\SHBB{2^{\bfj_1}}{r_3}\SHBB{1^{\bfi_1}}{4}\SHBB{1^{\bfi_2}}{4}\SHAA{2^{\bfj_2}}{3}}{m^3\SHBB{3}{r_3}}
 	\Big]		\mathfrak{H}_{-h_3,h_4}
 	\nonumber
 	\\
 	-&&\Big[\frac{\SHBB{1^{\bfi_1}}{r_4}\SHAB{3}{2}{r_3} \SHAB{2^{\bfj_1}}{3}{2^{\bfj_2}}\SHAA{1^{\bfi_2}}{4}}{m^4\SHBB{3}{r_3}\SHBB{4}{r_4}}+
 	\frac{\SHBB{2^{\bfj_1}}{r_3}\SHAA{1^{\bfi_1}}{4}\SHAA{1^{\bfi_2}}{4}\SHAA{2^{\bfj_2}}{3}}{m^3\SHBB{3}{r_3}}
 	\Big]		\mathfrak{H}_{-h_3,-h_4}
 	\nonumber
 \end{eqnarray}
 Adding this contact term makes the scattering amplitude gauge invariant. The expression above is not $1\leftrightarrow 2$ symmetric. This is simply due to the fact that we have written $\mathcal{G}_{\Upsilon^{(t)}}$. The full expression also involves $\mathcal{G}_{\Upsilon^{(u)}}$ which can be obtained from $\mathcal{G}_{\Upsilon^{(t)}}$ by $1\leftrightarrow 2$ exchange. The full expression of \eqref{emweinbergminimal52.1} is $1\leftrightarrow 2 $ symmetric.

{In the literature, one often makes the choice for reference spinors as $r_3=4$ \& $r_4=3$, and proceed with the computation. The above exercise shows that this cannot be done unless the amplitude is made manifestly gauge invariant by adding additional contact terms. As an example, in \cite{Arkani-Hamed:2017jhn} the authors compute the tree-level Compton amplitude such that at the poles it factorizes into a product of two on-shell three-point amplitudes which are manifestly invariant of the reference spinors. In our case, the contributions from the exchange diagrams are not manifestly independent of the reference spinors (in other words gauge invariant). As a consequence there are additional contact terms \eqref{emweinbergminimal71} which do not vanish for such a choice of the reference spinors\footnote{We thank the referee and Julio Parra Martinez for clarifying this point.}.}

 \subsection{Manifest gauge invariance}
 \label{subsec:krsfourpointIII}
 
 In the previous subsection, we showed that it is possible to find an analytic contact term such that when it is added to the contributions from the exchange Feynman diagram, it makes the full answer gauge invariant. So, the total contribution does not depend on a choice for the two reference vectors. In fact, it is possible to do something better. We show that after adding the contact term and applying a few Schouten rules, no term in the amplitude depends on the $r_{3}$ and $r_{4}$ and thus it is {\it manifestly independent of the choice of the reference vectors}. In that form, the amplitude is manifestly gauge-invariant. 
 
 Let's consider the contribution proportional to $\Omega(\mathcal{S})$ given in \eqref{emweinbergminimal64}. For simplicity, we consider $h_3=1=h_4$ and demonstrate it explicitly. In that case, the quantity in the parenthesis is given by 
 \begin{equation}
 	\frac{1}{\langle 3\,r_3\rangle\langle 4\,r_4\rangle}\Bigg(	\langle r_4|1|4]\langle r_3|2|3](u+m^2)
 	+
 	\langle r_4|2|4]\langle r_3|1|3](t+m^2)
 	+\langle r_3\,r_4\rangle[34](u+m^2)(t+m^2)\Bigg)
 	\label{emweinbergminimal101}
 \end{equation}
 We apply the following Schouten identity (we remind the reader that $i,j$ are index for massless legs and $a,b$ are indices for massive legs)
 \begin{equation}
 	\langle r_i|a|i]=\frac{\langle i\,r_i\rangle \langle j|a|i]-\langle j\,r_i\rangle\langle i|a|i]}{\langle ij\rangle}
 	\qquad,\qquad
 	j\ne i	
 	\label{emweinbergminimal102}
 \end{equation}
 We can see that there is an apparent pole in this expression. But this pole is fictitious since it actually cancels between the numerator and denominator. We will see that there is no such fictitious pole in the final expression. 
 
 After applying this (along with a few momentum conservation rules), the expression in \eqref{emweinbergminimal101} simplifies to 
 \begin{equation}
 	-m^2[34]^2
 	\label{emweinbergminimal103}
 \end{equation}
 For terms involving $\Upsilon^{(t)}(\mathcal{S})$ and $\Upsilon^{(u)}(\mathcal{S})$, the same method is applicable. We just need to include one identity along with \eqref{emweinbergminimal102}.
 \begin{equation}
 	\langle a\,r_i\rangle =\frac{\langle a\,3\rangle\langle r_i\,4\rangle-\langle a\,4\rangle\langle r_i\,3\rangle}{\langle 34\rangle}
 \end{equation}
 The final manifestly gauge invariant form of the amplitude $\mathcal{M}$, defined in \eqref{emweinbergminimal52.2}, for the IR-minimal three point vertex\footnote{See \eqref{emweinbergminimalthreepoint1} for the expression. This three point function is different from the three point function proposed in \cite{Arkani-Hamed:2017jhn} which is referred to as "minimal" by the authors of \cite{Arkani-Hamed:2017jhn}. Please see the first item of subsection \ref{subsecmainresult} for a comparison.} is given by 
 \begin{align}
 	\mathcal{M}=\tgem^2\left[\frac{m^2\Omega(\mathcal{S})}{(t+m^2)(u+m^2)}\mathfrak{G}_{\Omega}\,
 	+\frac{\Upsilon^{(t)}(\mathcal{S})}{(t+m^2)}\mathfrak{G}_{\Upsilon_t}
 	+\frac{\Upsilon^{(u)}(\mathcal{S})}{(u+m^2)}\mathfrak{G}_{\Upsilon_u}\right]\label{photonfinalanswer}
 \end{align}
 where,
 \small
 \begin{align}
 	\mathfrak{G}_\Omega(h_3,h_4) &= -\frac{\SHBB{3}{4}^2}{m^2}\mathfrak{H}_{h_3,h_4}
 	+\frac{\SHAB{3}{1}{4}^2}{m^4}\mathfrak{H}_{-h_3,h_4}
 	+\frac{\SHAB{4}{1}{3}^2}{m^4}\mathfrak{H}_{h_3,-h_4}
 	-\frac{\SHAA{3}{4}^2}{m^2}\mathfrak{H}_{-h_3,-h_4} \label{gluoncomp76}\\
 	\nonumber\\
 	\mathfrak{G}_{\Upsilon_t}(h_3,h_4) &= \frac{\SHBB{1^{\bfi_1}}{4}^2\SHBB{2^{\bfj_1}}{3}^2}{m^4}\mathfrak{H}_{h_3,h_4}
 	- \frac{\SHBB{1^{\bfi_1}}{4}^2\SHAA{2^{\bfj_1}}{3}^2}{m^4}\mathfrak{H}_{-h_3,h_4}-\frac{\SHAA{1^{\bfi_1}}{4}^2\SHBB{2^{\bfj_1}}{3}^2}{m^4}\mathfrak{H}_{h_3,-h_4}
 	+\frac{\SHAA{1^{\bfi_1}}{4}^2 \SHAA{2^{\bfj_1}}{3}^2}{m^4} \mathfrak{H}_{-h_3,-h_4}
 	\label{gluoncomp77}
 \end{align}
 \normalsize The symbols $\Omega(\spin)$ and $\Upsilon^{(t)}$ are defined in \eqref{emweinbergminimal15.1} and \eqref{emweinbergminimal15.2}. The $u$-channel factors $\Upsilon^{(u)}$ and $\mathfrak{G}_{\Upsilon_u}$ are just $ 1 \leftrightarrow 2 $ exchange of $\Upsilon^{(t)}$ and $\mathfrak{G}_{\Upsilon_t}$ respectively. In section \ref{sec:Comparisonwithliterature}, we will verify these results using the amplitude factorization method used in \cite{Arkani-Hamed:2017jhn,Chung:2018kqs}.  
{\paragraph{Comparison with the Lorentz covariant answer :}The gauge invariant IR minimal photon Compton amplitude for any spin was also computed in our previous paper \cite{Kumar:2025juz} in the Lorentz covariant notation in any $D$ spacetime dimension. In $D=3+1$ spacetime, the amplitude was given by
 \small 
 \begin{align}
\gem^2\left[-\frac{m^2\Omega(\spin)_{\text{Lor}}}{(t+m^2)(u+m^2)}\frac{2\left(k_1 \mathcal{W}^{(34)}k_1\right)}{m^4}+\frac{ \Upsilon^{(t)}(\spin)_{\text{Lor}}}{(t+m^2)}\frac{\left(k_1\mathcal{B}^{(4)}\zeta_1\right)\left(k_2\mathcal{B}^{(3)}\zeta_2\right)}{m^4}+\frac{ \Upsilon^{(u)}(\spin)_{\text{Lor}}}{(u+m^2)}\frac{\left(k_1\mathcal{B}^{(3)}\zeta_1\right)\left(k_2\mathcal{B}^{(4)}\zeta_2\right)}{m^4}\right]
 \end{align}
 \normalsize
 where the polynomials in the Lorentz notation are given by
 \begin{align}
 	\mathcal{B}^{(i)}_{\mu\nu}&=k_{i\mu}\varepsilon_{i\nu}-k_{i\nu}\varepsilon_{i\mu}\qquad,\qquad \mathcal{W}^{(34)}_{\mu\nu}=\mathcal{B}^{(3)}_{\mu\rho}\mathcal{B}^{(4)}_{\nu}{}^{\rho},\\
\Omega(\spin)_{\text{Lor}}&=(\zeta_1\cdot \zeta_2)^{\widehat \otimes_\spin}\quad,\quad\Upsilon^{(u)}(\spin)_{\text{Lor}}=\left.\Upsilon^{(t)}(\spin)_{\text{Lor}}\right|_{1\leftrightarrow2}, \\
\Upsilon^{(t)}(\spin)_{\text{Lor}}&=\sum_{b=0}^{\mathcal{S}-1}(-1)^{\mathcal{S}-b}\, \tilde B(\mathcal{S},b)
\left( \frac{\left(k_1\mathcal{B}^{(4)}\zeta_1\right)\left(k_2\mathcal{B}^{(3)}\zeta_2\right)}{m^4}\right)^{\widehat \otimes_{\mathcal{S}-b-1}}(\zeta_1\cdot \zeta_2)^{\widehat \otimes_b}
 \end{align}
 \normalsize
It is easy to check that the SH answer completely agrees with the Lorentz covariant answer by substituting the following conversions from Lorentz to SH
\begin{align}
	\gem=\sqrt{2}\tgem\quad&,\quad \Omega(\spin)_{\text{Lor}}=\Omega(\spin),\\
\left(k_1 \mathcal{W}^{(34)}k_1\right)=-\frac{m^4}{4}\mathfrak{G}_\Omega(h_3,h_4)\quad&,\quad  \Upsilon^{(t)}(\spin)_{\text{Lor}}= 2 \Upsilon^{(t)}(\spin),\\
\left(k_1\mathcal{B}^{(4)}\zeta_1\right)\left(k_2\mathcal{B}^{(3)}\zeta_2\right)=\frac{m^4}{4}\mathfrak{G}_{\Upsilon_t}(h_3,h_4)
\quad&,\quad \Upsilon^{(u)}(\spin)_{\text{Lor}}= 2 \Upsilon^{(u)}(\spin)
\end{align}}
 \paragraph{A few important features :}Here, we would like to emphasise a few important features
 \begin{enumerate}
 	\item We replace $r_3$ by $\tilde r_{3}$ (or $r_4$ by $\tilde r_{4}$), and compute the difference {\bf does not} vanish. However, there is no pole in the difference. More importantly, the poles in the difference cancel separately for the $t$ channel and for the $u$ channel. 
 	
 	\item Since the difference does not have a pole, it may be possible to cancel it by adding a local contact term. 
 	\item For the terms containing $\Omega(\mathcal{S})$, it is possible to find a contact term only after adding the contributions from both $t$-channel and $u$-channel. 
 	\item For the terms containing  $\Upsilon^{(t)}(\mathcal{S})$ or $\Upsilon^{(u)}(\mathcal{S})$, it is possible to find a contact term separately for $t$-channel (i.e. if we ad-hoc set $\Upsilon^{(u)}(\mathcal{S})$ to zero, it is still possible to find a contact term to make the amplitude gauge invariant).  
 \end{enumerate}

 \section{Compton amplitude with gluons}
 \label{sec:krsgluon}
 Until now, we have made the assumption that there are only two real (or one complex) massive spin $\mathcal{S}$ particles and only one massless spin one particle. However, it is possible that there are more particles: let's consider a slightly generalized situation. Let's say the number of real massive spin $\mathcal{S}$ particles is $N_F$, and the number of massless spin one particles is $N$. The choice of indices is the following
 \begin{equation}
 	\mathfrak{i}, \mathfrak{j}=1,\cdots, N_F
 	\qquad,\qquad
 	A,B =1, \cdots, N	
 	\label{gluoncomp1}
 \end{equation}
 So every massive particle carries index $\mathfrak{i}$, $\mathfrak{j}$, and every massless particle carries index $A$, $B$. Consider two massive particles labeled by index  $\mathfrak{i}$ and $\mathfrak{j}$ and a massless spin particle with label $A$; in this case, the three-point amplitude in \eqref{emweinbergminimalthreepoint1} modifies to
 \begin{equation}
 	\iimg \tgem \,T^A_{\mathfrak{i}\mathfrak{j}}\,  \mathcal{C}_{\mathcal{S}}^2\,
 	\frac{(x_{31}-x_{32})}{2}  \Big(\langle 1^{\bfi} 2^{\bfj} \rangle [1^{\bfi} 2^{\bfj}]\Big)^{\widehat \otimes_\mathcal{S} }
 	\qquad,\qquad
 	{\mathcal{C}_\mathcal{S}}^2=\frac{1}{(m^2)^{\mathcal{S}}}	
 	\label{gluoncomp3}
 \end{equation}
 Here $T^A_{\mathfrak{i}\mathfrak{j}}$ is a constant that depends on the labels of the three particles. It is known that for massive particles with spin 0, 1/2, 1, these interactions are highly constrained by unitarity (We encourage readers to check  Appendix B of \cite{Quevedo:2024kmy} for a clear exposition of this idea). The idea is generalized here, and we can illustrate it using the results from the previous section. In this case, \eqref{emweinbergminimal12.1} modifies to 
 \begin{equation}
 	\widetilde{\mathcal{M}}^{A_3A_4}_{\mathfrak{i}_2\mathfrak{i}_1}=
 	(\tgem)^2	(T^{A_3}T^{A_4})_{\mathfrak{i}_2\mathfrak{i}_1}\frac{\mathcal{F}^{(t)}(h_3,h_4)\mathcal{N}^{(t)}(\mathcal{S})	}{t+m^2}
 	+(\tgem)^2	(T^{A_4}T^{A_3})_{\mathfrak{i}_2\mathfrak{i}_1}\frac{\mathcal{F}^{(u)}(h_3,h_4)\mathcal{N}^{(u)}(\mathcal{S})	}{u+m^2}
 	\label{gluoncomp4}
 \end{equation} 
 $A_3 (A_4)$ is the flavour index\footnote{In the present context, the choice between flavour and colour depends on the perspective one wishes to emphasize. We begin with $N_F$ massive particles and $N$ massless particles, and the main conclusion is that gauge invariance is forced upon us if we require the Compton amplitude to be unitary. The term ``colour" is more commonly used in the Lagrangian approach, where one starts from a gauge-covariant derivative. Since our starting point is different, we find the term flavour to be more appropriate for the viewpoint adopted here, even though the final analysis shows that the index ultimately plays the role of a colour index.} for massless particle 3(4) and $\mathfrak{i}_1$($\mathfrak{i}_2$) is the flavour index for particle
 1 (2). 
 \begin{equation}
 	(T^{A_3}T^{A_4})_{\mathfrak{i}_2\mathfrak{i}_1}
 	=\sum_{\mathfrak{i}=1}^{ N_F}(T^{A_3})_{\mathfrak{i}_2\mathfrak{i}}(T^{A_4})_{\mathfrak{i}\mathfrak{i}_1}
 \end{equation}
 We can consider $T$s as matrices in the $\mathfrak{i}$ index, and then the above expression is simply matrix multiplication. We introduce a short-hand notation at this point. The usefulness of it, other than compactness of expressions, will be clear become clear later(in section \ref{sec:gluoncont}),
 \begin{equation}
 	\label{eq:Cdefs}
 	\mathrm{C}_t \equiv -T^{A_3}T^{A_4}, \qquad \mathrm{C}_u \equiv T^{A_4}T^{A_3} \quad\&\quad \mathrm{C}_s \equiv [T^{A_3}, T^{A_4}] =  \iimg f^{A_3A_4B}T^{B}  
 \end{equation}
 Then the exchange amplitude can be written as,
 \begin{align}
 	\widetilde{\mathcal{M}}^{(\mathrm{gluon})}
 	&=\tgem^2	\bigg[-\mathrm{C}_t\frac{\mathcal{F}^{(t)}\mathcal{N}^{(t)}(\mathcal{S})	}{t+m^2}
 	+\mathrm{C}_u\frac{\mathcal{F}^{(u)}\mathcal{N}^{(u)}(\mathcal{S})	}{u+m^2}\bigg]\nonumber\\
 	&=\tgem^2	\Bigg[
 	\bigg\{-\frac{\mathrm{C}_t\mathcal{F}^{(t)}}{t+m^2}
 	+\frac{\mathrm{C}_u\mathcal{F}^{(u)}}{u+m^2}\bigg\} \Omega(\spin)
 	-\mathrm{C}_t\frac{\langle 1^{\bfi_1}| 4 |1^{\bfi_{2}}]\langle 2^{\bfj_1}| 3 |2^{\bfj_{2}}]\mathcal{F}^{(t)}}{m^4(t+m^2)} \Upsilon^{(t)}(\mathcal{S})
 	+\langle 1\leftrightarrow 2 \rangle \Bigg]
 	\label{gluoncomp5}
 \end{align} 
 where we have expanded the $\mathcal{N}$s using \eqref{emweinbergminimal21}. It is easy to guess that the gauge non invariance of the $\Upsilon^{(t)}$(and $\Upsilon^{(u)}$) piece would not be different from the case of photon. It can also be easily fixed by almost the same expression as the case of photon. The only difference is that there would be a factor of $\mathrm{C}_s$ multiplied. The contact term is for it is given by,
 \begin{equation}
 	\label{eq:upsiloncont}
 	-\mathrm{C}_t\frac{\mathcal{G}_{\Upsilon_t}}{m^2}\Upsilon^{(t)}(\mathcal{S}) + \mathrm{C}_u\frac{\mathcal{G}_{\Upsilon_u}}{m^2}\Upsilon^{(u)}(\mathcal{S})
 \end{equation}
 
 Let us return to fix the gauge non-invariance of the $\Omega(\spin)$ piece in \eqref{gluoncomp5}. Unlike the case of photon, the two terms inside the curly braces in \eqref{gluoncomp5} can not be added up after changing $\mathcal{F}^{(ch)}$ to $\Delta\mathcal{F}^{(ch)}$. Fortunately, gluon has one more differing feature from photon; unlike photon, there exists three gluon vertex. This enables the contribution of s-channel also in the Compton amplitude. We will see in the next section, how s-channel amplitude helps in restoring gauge invariance. 
 
 \subsection{s-channel amplitude}
 \begin{figure}[H]
 	\begin{center}
 		\begin{tikzpicture}[line width=1.5 pt, scale=1.3]
 			
 			\begin{scope}[shift={(0,0)}]	
 				
 				
 				\treelevelthreepoint{gluon}{gluon}{gluon}{0}
 				\node at (-.75-.1,1.5) {$B$}; 
 				\node at (-.75-.1,-1.5) {$C$};
 				\node at (1.7,0) {$A$};
 				
 				
 			\end{scope}		
 		\end{tikzpicture} 
 	\end{center}
 	\caption{Gluon three-point vertex}
 	\label{fig:gluonthreepoint}
 \end{figure}
 Three-point vertex of three gluons labeled by $A$, $B$ and $C$  is given by, 
 \begin{equation}
 	f^{ABC}\frac{\lambda}{8m} \Big( k_{23}^{\dot{\alpha}_1\alpha_1} \epsilon^{\alpha_2 \alpha_3} \epsilon^{\dot\alpha_2 \dot\alpha_3} +k_{31}^{\dot{\alpha}_2\alpha_2} \epsilon^{\alpha_3 \alpha_1} \epsilon^{\dot\alpha_3 \dot\alpha_1} +k_{12}^{\dot{\alpha}_3\alpha_3} \epsilon^{\alpha_1 \alpha_2} \epsilon^{\dot\alpha_1 \dot\alpha_2}\Big)
 	\label{gluoncomp23}
 \end{equation}
 $f^{ABC}$ is a constant that depends on the three massless flavours; it must be anti-symmetric in any two indices so that Bose statistics of the gluons are obeyed. $\lambda$ is the coupling coefficient which needs to be determined by demanding gauge invariance. For simplicity of writing, we have stripped off the massless polarization, which can be given in terms of reference vectors as defined in \eqref{shbasics15}. One can check that, after putting the polarizations, it is gauge invariant (i.e. independent of the choice of the reference vectors). The factor of $\frac{1}{8}$ is kept such that the coupling matches with the following Lorentz expression 
 \begin{equation}
 	f^{ABC}\frac{\lambda}{2m} \bigg[\spp{\varepsilon_1}{k_{23}}\spp{\varepsilon_2}{\varepsilon_3}+\spp{\varepsilon_2}{k_{31}}\spp{\varepsilon_3}{\varepsilon_1}+\spp{\varepsilon_3}{k_{12}}\spp{\varepsilon_1}{\varepsilon_2}\bigg]
 \end{equation}
 The gluon propagator in the Feynman gauge is given by,
 \begin{equation}
 	\frac{-2\iimg \epsilon^{\alpha_1\beta_1} \epsilon^{\dot\alpha_2\dot\beta_2}}{p^2- \iimg\epsilon} \delta_{AB}
 	\label{gluoncomp25}
 \end{equation}
 where is $p$ is momenta.
 \begin{figure}[t]
 	\begin{center}
 		\begin{tikzpicture}[line width=1.5 pt, scale=1.3]
 			
 			\begin{scope}[shift={(0,0)}]	
 				
 				\treefourpointexchange{chspinbar}{chspin}{gluon}{gluon}{gluon}{-90}	
 				
 				\labeltreefourpointexchange{ $k_1$ }{ $k_2$ }{ $k_3$ }{ $k_4$ }{}{-90} 
 				
 				\draw (0,-2.5) node {$s$-channel exchange due to gluon exchange};
 			\end{scope}		
 		\end{tikzpicture} 
 	\end{center}
 	\caption{Higher-spin Compton: contribution from gluon exchange }
 	\label{fig:gluonschannel}
 \end{figure}
 The $s$-channel amplitude (see Fig \eqref{fig:gluonschannel}) is obtained sandwiching the propagator with the gluon self coupling \eqref{gluoncomp23} and the $\spin$-$\spin$-gluon minimal vertex \eqref{gluoncomp3} with the gluons stripped\footnote{Stripping the gluon leg requires substituting $x_{31}-x_{32}$ with $ \frac{1}{\sqrt{2}}k_{12}^{\rho\dot{\rho}}$.}. The $s$ channel amplitude is given by $\iimg \widetilde{\mathcal{M}}_s$ where,
 \begin{equation}
 	\label{eq:samp}
 	\widetilde{\mathcal{M}}_s^{(\mathrm{gluon})} =\mathrm{C}_s \tgem \frac{\lambda}{2\sqrt{2}} \frac{\mathcal{F}^{(s)}}{s}\Omega(\spin)
 \end{equation}
 where 
 \begin{align}
 	\label{eq:deffs}
 	\mathcal{F}^{(s)} \equiv \mathcal{F}^{(t)}-\mathcal{F}^{(u)}+\frac{t-u}{2 m^2}\mathcal{G}_\Omega
 \end{align}
 where $\mathcal{F}^{(t)}$ is given in \eqref{emweinbergminimal22} and $\mathcal{F}^{(u)}$ is just $\langle 1 \leftrightarrow 2 \rangle$ exchange of it. $\mathcal{G}_\Omega$ is defined in \eqref{emweinbergminimal63}. Now let us check what is the gauge non-invariant part of this amplitude.
 
 \subsection{Restoring gauge invariance and flavour kinematics duality}
 \label{sec:gluoncont}
 Now that we have the s channel amplitude, let us return to restoring the gauge invariance for the $\Omega$ piece in \eqref{gluoncomp5}. After including the $s$-channel contribution, we can write
 \begin{align}
 	\widetilde{\mathcal{M}}_\Omega^{(\mathrm{gluon})}&\equiv\tgem\bigg\{-\tgem\frac{\mathrm{C}_t\mathcal{F}^{(t)}}{t+m^2}
 	+\tgem\frac{\mathrm{C}_u\mathcal{F}^{(u)}}{u+m^2}
 	+\frac{\lambda}{2\sqrt{2}} \frac{\mathrm{C}_s\mathcal{F}^{(s)}}{s}\bigg \}\Omega(\spin)
 	\label{Mgluonomega}
 \end{align}
 Now note that,
 \begin{align}
 	\frac{\Delta_3{\mathcal{F}^{(t)}}}{t+m^2} =  \frac{x_{41}}{m^2}A_3& \qquad\qquad\qquad \frac{\Delta_4{\mathcal{F}^{(t)}}}{t+m^2} =  \frac{x_{32}}{m^2}A_4\\
 	\frac{\Delta_3{\mathcal{F}^{(u)}}}{u+m^2} = \frac{x_{42}}{m^2}A_3& \qquad\qquad\qquad \frac{\Delta_4{\mathcal{F}^{(u)}}}{u+m^2} = \frac{x_{31}}{m^2}A_4\\
 	\frac{\Delta_3{\mathcal{F}^{(s)}}}{s} = -\frac{x_{41}-x_{42}}{2m^2} A_3&
 	\qquad\qquad\qquad \frac{\Delta_4{\mathcal{F}^{(s)}}}{s} = -\frac{x_{32}-x_{31}}{2m^2} A_4
 \end{align}
 where $A_i$ is defined in \eqref{eq:Adef}. Then,
 \begin{equation}
 	\Delta_3 \widetilde{\mathcal{M}}_\Omega^{(\mathrm{gluon})}
 	= \tgem \frac{A_3}{m^2}\bigg\{ \bigg(-\tgem \mathrm{C}_{t}-\frac{\lambda}{4\sqrt{2}}\mathrm{C}_s\bigg)x_{41}+ \bigg(\tgem \mathrm{C}_u+\frac{\lambda}{4\sqrt{2}}\mathrm{C}_s\bigg)x_{42}\bigg\}\Omega(\spin)
 \end{equation}
 We have to either make the above term zero or bring into a form by putting relation between $\mathrm{C}_s$ and value of $\lambda$ such that we can find a local term whose gauge non invariance of the leg 3 is same as above. We already know a contact term  $\mathcal{G}_{\Omega}$ from section \ref{sec:findingGomega} whose gauge non-invariant piece is  $\sim A_3 (x_{41}+x_{42})=-A_3x_{43}$. To bring the above expression to this form, we need to set,
 \begin{equation}
 	\label{eq:sumC0}
 	\lambda = \pm2 \sqrt{2}\tgem \qquad \& \quad \mathrm{C}_t+\mathrm{C}_u\pm\mathrm{C}_s=0
 \end{equation}
 Finally, the contact term needed for gauge invariance which needs to be added to \eqref{Mgluonomega} is,
 \begin{equation}
 	\frac{\tgem^2}{2m^2}\left( \mathrm{C}_u-  \mathrm{C}_t\right)\mathcal{G}_\Omega \Omega(\spin)
 \end{equation}
 After adding it to \eqref{Mgluonomega}, the final gauge invariant expression is
 \begin{equation}
 	\label{eq:MgluonomegaGI}
 	\mathcal{M}_\Omega^{(\mathrm{gluon})} = \tgem^2\bigg\{\frac{\mathrm{C}_t(-\mathcal{F}^{(t)}-\frac{t+m^2}{2m^2}\mathcal{G}_\Omega)}{t+m^2}
 	+\frac{\mathrm{C}_u(\mathcal{F}^{(u)}+\frac{u+m^2}{2m^2}\mathcal{G}_\Omega)}{u+m^2}
 	\pm\frac{\mathrm{C}_s\mathcal{F}^{(s)}}{s}\bigg \}\Omega(\spin)
 \end{equation}
 This if written in the the follwing form,
 \begin{equation}
 	\mathcal{M}_\Omega^{(\mathrm{gluon})} = \tgem^2\bigg\{\frac{\mathrm{C}_t\eta_t}{t+m^2}
 	+\frac{\mathrm{C}_u\eta_u}{u+m^2}
 	+\frac{\mathrm{C}_s\eta_s}{s}\bigg \}\Omega(\spin)
 \end{equation}
 it gives, 
 \begin{equation}
 	\eta_t +\eta_u \pm\eta_s =0
 \end{equation}
 where we have used the definition in \eqref{eq:deffs}. In this expression, $\mathrm{C}_s$ are the flavour part which just have the gauge symmetry generators (defined in \eqref{eq:Cdefs}), while $\eta$s are the kinematics part which just have the polarizations and momenta. Both of them follow the same form of relation(above and \eqref{eq:sumC0}). This is flavour kinematics duality. \cite{Bern:2010ue} hypothesizes and shows for considerable amount of cases that if flavour part and kinematics part of a gauge field theory follow the same relation in a gauge theory amplitude, we can directly use it to figure out the amplitude for the process where external gauge legs are replaced with gravitons. We just need to replace the flavour parts in the amplitude($\mathrm{C}_{ch}$) with the corresponding kinematics part($\eta_{ch}$). Hence if the flavour kinematics duality is valid for higher-spins, we can easily predict the omega piece of gravity amplitude as, 
 \begin{equation}
 	\mathcal{M}_\Omega^{(\mathrm{graviton})} \sim \bigg\{\frac{\eta_t^2}{t+m^2}
 	+\frac{\eta_u^2}{u+m^2}
 	+\frac{\eta_s^2}{s}\bigg \}\Omega(\spin)
 \end{equation}
 
 \subsection{Manifestly gauge invariant form}
 Since the $\Omega$, and $\Upsilon$ pieces are separately gauge invariant, it is possible to write their manifestly gauge invariant form separately. For the $\Upsilon$ pieces we already know the gauge invariant answer from section \ref{subsec:krsfourpointIII}. $\Omega$ piece is new. Let us start from gauge invariant $\Omega$ piece in \eqref{eq:MgluonomegaGI}. We need to invoke \eqref{eq:sumC0}, because it is gauge invariant only when this relation is satisfied. We replace $\mathrm{C}_s$ using this. Then the expression becomes,
 \begin{equation}
 	\mathcal{M}_\Omega^{(\mathrm{gluon})} = \tgem^2\bigg\{-\mathrm{C}_t\bigg(\frac{\mathcal{F}^{(t)}}{t+m^2}+\frac{\mathcal{F}^{(s)}}{s}+\frac{\mathcal{G}_\Omega}{2m^2}\bigg)
 	+\mathrm{C}_u\bigg(\frac{\mathcal{F}^{(u)}}{u+m^2}-\frac{\mathcal{F}^{(s)}}{s}+\frac{\mathcal{G}_\Omega}{2m^2}\bigg)
 	\bigg \}\Omega(\spin)
 \end{equation}
As an example, for $h_3=1=h_4$ the terms in the round brackets becomes proportional to \eqref{emweinbergminimal101} which can then be written as \eqref{emweinbergminimal103} in a manifestly gauge invariant form using Schouten identities. Similar results would follow for other helicities. Finally, we can write the gauge invariant amplitude also including the $\Upsilon_t$ and $\Upsilon_u$ parts as
 \begin{equation}
 	\frac{\mathcal{M}^{(\mathrm{gluon})}}{\tgem^2}=\frac{m^2}{s}\bigg(\frac{\mathrm{C}_t}{t+m^2}-\frac{\mathrm{C}_u}{u+m^2}\bigg)\Omega(\mathcal{S})\mathfrak{G}_\Omega\
 	-\frac{\mathrm{C}_t}{t+m^2}\Upsilon_t(\mathcal{S})\mathfrak{G}_{\Upsilon_t}
 	+\frac{\mathrm{C}_u}{u+m^2}\Upsilon_u(\mathcal{S})\mathfrak{G}_{\Upsilon_u}
 	\label{eq:gluonfinalamp}
 \end{equation}
 where $\mathfrak{G}_{\Omega}$ and $\mathfrak{G}_{\Upsilon_t}$ are the same scalars which appeared in the gauge invariant amplitude for photon. These are defined in \eqref{gluoncomp76} and \eqref{gluoncomp77}. Note that setting $C_t=-C_u=-1$ gives back the photon answer.
 

 \section{Ghosts for $\xi=1$}
 \label{sec:krsphotonghost}
 
 In \cite{Kumar:2025juz}, the authors contemplated whether there is an analogue of $R_\xi$ gauge in any theory of massive higher-spin particles. Since there is no suitable Lagrangian, one can ask the questions directly at the level of on-shell amplitudes. In that, one can make an ansatz for the $R_\xi$ gauge. In  \cite{Kumar:2025juz}, the authors made an ansatz which becomes \eqref{SHpropagator12} in the SH formalism. All the calculations we did in previous sections have been based on the massive spin $\mathcal{S}$ particle in unitary Gauge ($\xi=\infty$). What happens for other values of $\xi$? In \cite{Kumar:2025juz}, the authors have given the results for $\xi=1$. In there, they needed to add a tower of ghosts\footnote{Here, the ghost means unphysical particles of the same statistics; these are not Fadeev-Popov ghosts. In the context of Higgesed gauge theory, these are the ``would-be goldstone" bosons.} to make the amplitude unitary. We will follow a similar procedure for $\xi=1$ and give the ghost couplings in SH formalism. For this case, only one term remains in the expression of $\Theta$ in \eqref{SHpropagator12}. The computation simplifies hugely. $\big(1\odot_t 1\big)$ and  $\big(2\odot_t 2\big)$ would be zero and $\big(1\odot_t 2\big)$ would be just the first term of \eqref{emweinbergminimal4.3}. Then the $\mathcal{N}_t(\mathcal{S})$ would just have a single term $\Omega(\mathcal{S})$. The final gauge invariant amplitude is given by, 
 
 \begin{equation}
 	\mathcal{M}_{\xi=1}=\tgem^2	\Bigg[\frac{\mathcal{F}^{(t)}(h_3,h_4)	}{t+m^2}
 	+\frac{\mathcal{F}^{(u)}(h_3,h_4)}{u+m^2}+\frac{\mathcal{G}_\Omega(h_3,h_4)}{m^2}\Bigg]\Omega(\mathcal{S})=\tgem^2 \frac{m^2\mathfrak{G}_{\Omega}(h_3,h_4)\Omega(\spin)}{(t+m^2)(u+m^2)}
 	\label{emweinbergminimal12.1xi1}
 \end{equation}
 One could check using Cutkowsky rules that the above amplitude is not unitary. Just like the case of \cite{Kumar:2025juz}, to make it unitary, we need to add ghost exchanges details of which could be determined based on the difference of amplitude in unitary gauge \eqref{emweinbergminimal71} and above amplitude. It is easy to see that only the $\Upsilon$ part has to be contributed by the tower of ghosts. These ghosts would be of spin ($\mathcal{S}_{g}$) ranging from $0$ to $\mathcal{S}-1$. One can calculate the required massive spin-$\mathcal{S}$, spin-$\mathcal{S}_{g}$ ghost, photon coupling constant as,
 \begin{equation}
 	\iimg \left(\frac{1}{m}\right)^{\mathcal{S}-s}(\sqrt{2})^{s}  \sqrt{\tilde{B}(\mathcal{S}, \mathcal{S}_{g})}\ x\ \Big(
 	\SHAA{1^{\bfi_j}}{2^{\bfj_j}}\SHBB{1^{\bfi_{j+1}}}{2^{\bfj_{j+1}}}
 	\Big)^{\widehat \otimes_{\mathcal{S}_{g}}}
 	\Big(\SHAB{1^{\bfi_j}}{3}{1^{\bfi_{j+1}}}\Big)^{\widehat\otimes_{\mathcal{S}-\mathcal{S}_{g}-1}}
 \end{equation}
 
\section{Comparison with the literature} \label{sec:Comparisonwithliterature}
In this section, we compare the results we have obtained with the methods existing literature. In \cite{Haddad:2023ylx}\footnote{We thank the referee for providing this reference.}, the authors computed the graviton Compton amplitudes for the most general three-point interactions using the BCFW recursion technique; the computations were restricted to sub-leading order in $\hbar$ expansion in order to match with the classical black-hole scattering. In other works \cite{Arkani-Hamed:2017jhn,Cangemi:2023ysz, Chung:2018kqs, Cangemi:2022bew}, the authors used the BCFW/amplitude factorization to compute the Compton amplitude for the AHH minimal three-point interaction. However, the spurious pole free Compton amplitude for any three-point interaction apart from the UV minimal at arbitrary orders of $\hbar$ is not available in the literature.

 In this section, we use the amplitude factorization approach and verify our results in \eqref{photonfinalanswer} with this method for any spin and helicities. The Feynman diagrammatic approach has both the advantages and disadvantages over the amplitude factorization approach. The key advantage is that the later leads to amplitudes with spurious poles which, as we will see, must be carefully removed by non-trivial expansions of the type \eqref{AHHnonlocalterm}. Whereas the Feynman diagrammatic approach is manifestly free of spurious poles which makes it more transparent at the level of analytic structure. On the other hand, unlike amplitude factorization approach the Feynman diagrammatic is not manifestly gauge invariant. As we have shown, extra contact terms are needed to preserve gauge invariance which are easily taken care by the algorithm we have provided. Beyond these, the Feynman diagrammatic approach can be easily extended to higher dimensions and arbitrary gauge choices, as we discussed in section \ref{sec:krsphotonghost} and was also explored in \cite{Kumar:2025juz}. This flexibility becomes crucial for higher-spin theories where computations in arbitrary gauge choices are relevant for UV completion.
\subsection{Photon Compton amplitude}
Consider the $t$-channel (figure \ref{fig:BCFWTCh}) for the photon Compton scattering with a cut on the internal propagator which puts it on-shell.
 \begin{figure}[H]
	\begin{center}
		\begin{tikzpicture}[line width=1.5 pt, scale=1.3]

			\begin{scope}[shift={(0,0)}, rotate=0]
				\draw[chspinbar] (0,0)--(1.5,0);   
				\draw[chspin][rotate=120] (0,0)--(1.5,0);   
				\draw[photon][rotate=-120] (0,0)--(1.5,0);   
				
				\draw[->] (.35,.3)--(1.3,.3);
				\draw[->][rotate=120] (1.5,.5)--(.75,.5);
				\draw[->][rotate=-120] (1.5,-.5)--(.75,-.5);

				\node at (-.75-.5,.4) {$k_2$}; 
				\node at (-.75-.5,-.4) {$k_3$};
				\node at (1.,.55) {$P$};
				
			\end{scope}	
				
			\begin{scope}[shift={(1.625,0)}, rotate=0]

				\draw[dotted] (0,1.35)--(0,-1.35);

			\end{scope}	
			
			\begin{scope}[shift={(3.25,0)}, rotate=180]
			\draw[chspin] (0,0)--(1.5,0);   
			\draw[photon][rotate=120] (0,0)--(1.5,0);   
			\draw[chspinbar][rotate=-120] (0,0)--(1.5,0);   
			
			\draw[->] (1.35,-.3)--(.35,-.3);
			\draw[->][rotate=120] (1.5,.5)--(.75,.5);
			\draw[->][rotate=-120] (1.5,-.5)--(.75,-.5);

			\node at (-.75-.5,.4) {$k_4$}; 
			\node at (-.75-.5,-.4) {$k_1$};
			\node at (1.,-.55) {$P$};
			
		\end{scope}	
		\end{tikzpicture} 
	\end{center}
	\caption{$t$-channel factorization}
	\label{fig:BCFWTCh}
\end{figure}

The residue at the $t=-m^2$ pole is given by the product of the left and the right three-point vertices 
\begin{align}
\left.\mathcal{M}(h_3,h_4)\right|_{t=-m^2}&=\mathcal{C}_\spin^4\tgem^2 \mathcal{F}^{(t)}(h_3,h_4)\left(\SHAA{P_{(J}}{\,1^{\bfi}}\SHBB{P_{K)}}{\,1^{\bfi}}\SHAA{2^{\bfj}}{\,P^{(J}}\SHBB{2^{\bfj}}{\,P^{K)}}\right)^{\hat{\otimes}_\spin}\label{BCFWtchresidue}
\end{align}
where $\mathcal{F}^{(t)}(h_3,h_4)=x_{32}(h_3)x_{41}(h_4)$ defined in \eqref{emweinbergminimal22}. Since the three-point vertices are on-shell gauge-invariant, we can choose the reference spinors $r_3=4$ and $r_4=3$ without worrying about the contact term completion.
\small
\begin{align}
	\left.	\mathcal{F}^{(t)}(h_3,h_4) \right|_{r_3=4,r_4=3}= 
	\frac{\langle 4|1|3]\langle 3|1|4][3\,4]^2 }{m^2s^2}&\mathfrak{H}_{h_3,h_4}
	-\frac{\langle 4|1|3]^2}{m^2s }\mathfrak{H}_{h_3,-h_4}
	-\frac{\langle 3|1|4]^2}{m^2s}\mathfrak{H}_{-h_3,h_4}
	+
	\frac{\langle 3|1|4] \langle 4|1|3] \langle3\, 4\rangle^2}{m^2s^2}\mathfrak{H}_{-h_3,-h_4}
\end{align}
\normalsize
The product $\langle 3|1|4]\langle 4|1|3]$ can also be written as $m^2s+(t+m^2)(u+m^2)$. At $t=-m^2$ pole, it is simply $\left. \langle 3|1|4]\langle 4|1|3]\right|_{t=-m^2}=m^2s$. Using this decomposition it is clear that at $t=-m^2$ (or $s=-(u+m^2)$) the residue has poles which can be interpreted as a pole from either the $s$ channel with a photon exchange or from $u$ channel with a massive exchange. As discussed in sec \ref{sec:CanotherExchanges?}, the three photon couplings vanish so the $s$ channel doesn't contribute. So the pole can only be interpreted as $u+m^2$. This is not a new feature but rather true for theories with $x$ factors i.e. theories where massive particles are coupled with massless particles (see \cite{Arkani-Hamed:2017jhn,Chung:2018kqs, Cangemi:2022bew,Cangemi:2023ysz} in case of the UV minimal amplitudes). In case of IR minimal, the prefactor can be written as
\begin{align}
	\left.	\mathcal{F}^{(t)}(h_3,h_4) \right|_{r_3=4,r_4=3, t=-m^2}=\frac{	m^2}{u+m^2} \mathfrak{G}_{\Omega}(h_3,h_4)\label{BCFWFfactor}
\end{align}
where $\mathfrak{G}_\Omega$ is defined in \eqref{gluoncomp76}.
The internal momenta $P$ satisfies the on-shell identities (see \cite{Arkani-Hamed:2017jhn} for more discussion)
\begin{align}
	P_{\alpha\dot{\alpha}}[4| ^{\alpha }=m x_{41}(+1) \langle 4| _{\dot{\alpha }}\quad,\quad P_{\alpha\dot{\alpha}}| 3\rangle ^{\dot{\alpha }}=-m x_{32}(-1)| 3]_{\alpha }\quad,\quad P_{\alpha\dot{\alpha}}P^{\dot{\alpha}\alpha}=2m^2
\end{align}
The solution to these on-shell conditions with reference to  \cite{Arkani-Hamed:2017jhn,Chung:2018kqs} is given by
\begin{align}
	P^{\dot{\alpha}\alpha}=\frac{m^2 [3| ^{\alpha } | 4\rangle ^{\dot{\alpha }}-\left[32^{J}\right] \left\langle
		1_{I}4\right\rangle  \left[1^{I}\right| ^{\alpha } \left|
		2_{J}\right\rangle {}^{\dot{\alpha }}}{\langle 4| 1| 3]}
\end{align}
The full expansion of the numerator $\left(\SHAA{P_{(J}}{\,1^{\bfi}}\SHBB{P_{K)}}{\,1^{\bfi}}\SHAA{2^{\bfj}}{\,P^{(J}}\SHBB{2^{\bfj}}{\,P^{K)}}\right)^{\hat{\otimes}_\spin}$ is given by the external higher-spin polarizations sandwiched with the unitary gauge higher-spin projector $\mathcal{P}_{(\mathcal{S})}^{\{\alpha,\dot \alpha\},\{\beta, \dot \beta \},}$ given in \eqref{SHpropagator3} where the spin $1$ projector $\Theta^{\alpha\dot{\alpha}\beta\dot{\beta}}(P)$ is given by
\begin{align}
\Theta^{\alpha\dot{\alpha}\beta\dot{\beta}}(P)=\sum \zeta(P)^{\dot{\alpha}\alpha}(\zeta(P)^{\dot{\beta}\beta})^*=-2\epsilon^{\alpha\beta} \epsilon^{\dot\alpha \dot\beta}  + \frac{P^{\dot\alpha \alpha} P^{\dot\beta \beta}}{m^2}
\end{align} 
where $\zeta(P)^{\dot{\alpha}\alpha}$ are the spin $1$ on-shell polarizations of the intermediate leg given by $\frac{\sqrt{2}}{m} |P^{J}\rangle^{\dot{\alpha}}[P^J|^{\alpha}$.
Let's denote any arbitrary scalar computed using the spin $1$ projector as
\begin{align}
	\big(X\odot_P Y\big)\equiv 	\langle X|^{(\bfi_1}_{\dot\alpha_1} |X]^{\bfi_2)}_{\alpha_1 } 
\Theta^{\alpha_1 \dot\alpha_1 \alpha_2  \dot\alpha_2}(P)	
\langle Y|^{(\bfj_1}_{\dot\alpha_2} |Y]^{\bfj_2)}_{\alpha_2}\quad,\quad 	\big(Y\odot_P X\big)=\big(X\odot_P Y\big)
\end{align}
Then the full numerator can be written as the following summation over $	\big(1\odot_P 1\big)$,$	\big(2\odot_P 2\big)$ and $	\big(1\odot_P 2\big)$\footnote{Note that it would be wrong to write $\left(\SHAA{P_{(J}}{\,1^{\bfi}}\SHBB{P_{K)}}{\,1^{\bfi}}\SHAA{2^{\bfj}}{\,P^{(J}}\SHBB{2^{\bfj}}{\,P^{K)}}\right)^{\hat{\otimes}_\spin}=\SHAB{1^{\bfi}}{P}{2^{\bfj}}^{\hat{\otimes}_\spin}\SHAB{2^{\bfj}}{P}{1^{\bfi}}^{\hat{\otimes}_\spin}$ because the massive Little group indices on the RHS are all symmetrized and so there are trace terms $\SHAB{1^{\bfi}}{P}{1^{\bfi}}\SHAB{2^{\bfj}}{P}{2^{\bfj}}$ with proper factors which are governed by \eqref{BCFWTChNumerator}. In case of AHH minimal amplitude, depending on th helicities, the numerator is as simple as $\SHAB{2^{\bfj}}{P}{1^{\bfi}}^{\hat{\otimes}_{2\spin}} $ or $\SHAA{1^{\bfi}}{2^{\bfj}}^{\hat{\otimes}_2\spin}$ because of the absence of trace terms.}.
\begin{align}
\left(\SHAA{P_{(J}}{\,1^{\bfi}}\SHBB{P_{K)}}{\,1^{\bfi}}\SHAA{2^{\bfj}}{\,P^{(J}}\SHBB{2^{\bfj}}{\,P^{K)}}\right)^{\hat{\otimes}_\spin}=\frac{1}{2^\spin \mathcal{C}_\spin^2}\sum_{a=0}^{\lfloor\frac{\spin}{2}\rfloor} \tilde{A}(\spin,a)	\big(1\odot_P 1\big)^{\hat{\otimes}_a}	\big(2\odot_P 2\big)^{\hat{\otimes}_a}	\big(1\odot_P 2\big)^{\hat{\otimes}_{\spin-2a}}\label{BCFWTChNumerator}
\end{align}
The factor of $\frac{1}{2^\spin \mathcal{C}_\spin^2}$ appears because there are $2\spin$ number of spin $1$ polarizations.
Note that even though the form of $P^{\alpha\dot{\alpha}}$ contains a spurious pole, the scalars which appear in \eqref{BCFWTChNumerator} turns out to have a local form at the $t=-m^2$ pole. It is easy to show that the scalars $\big(X\odot_P Y\big)$ are same as the scalars defined in \eqref{emweinbergminimal4.1}-\eqref{emweinbergminimal4.3}
 \begin{eqnarray}
	\big(X\odot_P Y\big)=\big(X\odot_t Y\big)\qquad\forall\,  X,Y\in \{1,2\}
\end{eqnarray}
This implies that, up to a factor of $\frac{1}{\mathcal{C}_\spin^4}$, the numerator \eqref{BCFWTChNumerator} is the same as the numerator that appears in the Feynman diagrammatic approach, defined as $\mathcal{N}^{(t)}(\spin)$ in \eqref{emweinbergminimal13}.
 \begin{align}
\left(\SHAA{P_{(J}}{\,1^{\bfi}}\SHBB{P_{K)}}{\,1^{\bfi}}\SHAA{2^{\bfj}}{\,P^{(J}}\SHBB{2^{\bfj}}{\,P^{K)}}\right)^{\hat{\otimes}_\spin}=\frac{1}{\mathcal{C}_\spin^4}\mathcal{N}^{(t)}(\spin)\label{BCFWTChNumerator2}
 \end{align}
By inserting \eqref{BCFWFfactor} and \eqref{BCFWTChNumerator2} in \eqref{BCFWtchresidue}, we obtain the $t$-channel gauge-invariant residue. Similarly, the $u$-channel residue can be computed. It is simply given by $1\leftrightarrow2$ exchange of the $t$-channel residue.
 \begin{align}
\left.\mathcal{M}\right|_{t=-m^2}&=\tgem^2 \frac{	m^2\mathfrak{G}_{\Omega}}{u+m^2} \mathcal{N}^{(t)}(\spin),\\
\left.\mathcal{M}\right|_{u=-m^2}&=\tgem^2 \frac{	m^2\mathfrak{G}_{\Omega}}{t+m^2} \mathcal{N}^{(u)}(\spin)
 \end{align}
 To construct the full amplitude using these residues, we first note that following \eqref{emweinbergminimal21} the numerator $\mathcal{N}^{(t)}(\spin)$ at $t=-m^2$ pole can be expanded as follows
\begin{align}
\left.\mathcal{N}^{(t)}(\spin)\right|_{t=-m^2}=\Omega(\spin)+\frac{(u+m^2)}{m^2}\frac{\mathfrak{G}_{\Upsilon_t}}{\mathfrak{G}_{\Omega}}\Upsilon^{(t)}(\spin) 
\end{align}
where $\mathfrak{G}_{\Upsilon_t}$ is defined in \eqref{gluoncomp77} and we have used the relation (see appendix \ref{app:GOmegarelation} for derivation)
 \begin{align}
 \left.\mathfrak{G}_{\Upsilon_t}(h_3,h_4)\right|_{t=-m^2}=\mathfrak{G}_{\Omega}(h_3,h_4)\frac{\SHAB{1^{\bfi}}{4}{1^{\bfi}}\SHAB{2^{\bfj}}{3}{2^{\bfj}} }{m^2(u+m^2)},\label{OmgeaUpsilontrelation}
\end{align}
 This implies that the $t=-m^2$ residue is a sum of two terms : the term proportional to $\Omega(\spin)$ which has a $u+m^2$ pole and the term proportional to $\Upsilon^{(t)}(\spin)$ which is completely local. The $u$-channel residue can be obtained by $1\leftrightarrow2$ exchange.
 \begin{align}
 	\left.\mathcal{M}\right|_{t=-m^2}&=\tgem^2 \left[\frac{m^2\mathfrak{G}_\Omega}{u+m^2}\Omega(\spin)+\mathfrak{G}_{\Upsilon_t}\Upsilon^{(t)}(\spin) \right],\label{Tchannelresidue}\\
 	\left.\mathcal{M}\right|_{u=-m^2}&=\tgem^2 \left[\frac{m^2\mathfrak{G}_\Omega}{t+m^2}\Omega(\spin)+\mathfrak{G}_{\Upsilon_u}\Upsilon^{(u)}(\spin) \right],\label{UChannelresidue}\\
 	\left.\mathcal{M}\right|_{t=-m^2,u=-m^2}&=\tgem^2m^2 \mathfrak{G}_\Omega\Omega(\spin)\label{DoublePoleresidue}
 \end{align}
  Note that the term proportional to $\Omega(\spin)$ contributes to the double pole $t=-m^2$ and $u=-m^2$. Therefore, the full IR minimal amplitude is given by the sum of contributions from the $t+m^2$ and $u+m^2$ residues minus the contribution at the double pole which is common to both the channels.
 \begin{align}
	\mathcal{M}&=\frac{\left.\mathcal{M}\right|_{t=-m^2}}{t+m^2}+\frac{\left.\mathcal{M}\right|_{u=-m^2}}{u+m^2}-\frac{\left.\mathcal{M}\right|_{t=-m^2,u=-m^2}}{(t+m^2)(u+m^2)},\\
	&=\tgem^2\left[\frac{m^2\mathfrak{G}_\Omega}{(t+m^2)(u+m^2)}\Omega(\spin)+\frac{\mathfrak{G}_{\Upsilon_t}}{(t+m^2)}\Upsilon^{(t)}(\spin) +\frac{\mathfrak{G}_{\Upsilon_u}}{(u+m^2)}\Upsilon^{(u)}(\spin)\right]
\end{align}
which is same as the amplitude \eqref{photonfinalanswer} obtained from the Feynman diagrammatic approach for any helicities of the photons. Let's take this opportunity to compare this IR minimal amplitude with the the UV minimal amplitude computed in \cite{Arkani-Hamed:2017jhn,Chung:2018kqs} which for $h_3=+1,h_4=-1$ is given by
\begin{align}
\mathcal{M}^{(UV)}_{+-}&=\frac{\SHAB{4}{1}{3}^2}{m^2 (t+m^2)(u+m^2)}\left(\frac{\SHAA{2^{\bfj}}{4} \SHBB{1^{\bfi}}{3}+\SHAA{1^{\bfi}}{4} \SHBB{2^{\bfj}}{3}}{\SHAB{4}{1}{3}} \right)^{\hat{\otimes}_{2\spin}}
\end{align}
In this form, the amplitude scales like $(-1)^{2\spin}$ under the $1\leftrightarrow2$ exchange which reflects the spin statistics as a consequence of Poincare invariance and unitarity. Also it has the correct factorization at the poles $u+m^2$ and $t+m^2$. However, for $\spin\geq 2$ the expression has a spurious pole which in \cite{Arkani-Hamed:2017jhn,Chung:2018kqs} is resolved using the following decomposition
\small
\begin{align}
	\left.\frac{\SHAA{2^{\bfj}}{4} \SHBB{1^{\bfi}}{3}+\SHAA{1^{\bfi}}{4} \SHBB{2^{\bfj}}{3}}{\SHAB{4}{1}{3}} \right|_{t=-m^2}&=\frac{
		\left(-\left\langle 1^{\bfi}\right| 3| 2^{\bfj}]+\left\langle 2^{\bfj}\right| 3| 1^{\bfi}]+2 m \left[1^{\bfi}\,2^{\bfj}\right]\right)}{2 m^2}-\frac{(u+m^2) \left[2^{\bfj}\,3\right] \left\langle 1^{\bfi}\,4\right\rangle }{2 m^2 \langle 4| 1| 3]} \equiv A+B \label{AHHnonlocalterm}
\end{align}
\normalsize
The term $A$ is antisymmetric in $1\leftrightarrow2$ exchange and so will produce the correct residue at the $u+m^2$ pole. The term $B$ doesn't contribute at the $u+m^2$ pole and it's spurious pole can be removed using the following 
\begin{align}
	\left.B \right|_{t=-m^2}=-\frac{[3\,4] \left\langle 2^{\bfj}\,3\right\rangle  \left\langle 1^{\bfi}\,4\right\rangle }{2 m^3}
\end{align}
After incorporating this, the UV minimal amplitude for any spin can be written as a sum of three terms : a double pole term proportional to $A^{2\spin}$ which scales as $(-1)^{2\spin}$ under $1\leftrightarrow2$ exchange, a term which is a function of $A$ and $B$ which contributes only at $t+m^2$ pole and a similar term which contributes only at the $u+m^2$ pole. The IR minimal amplitude we have computed has the similar behavior : the term proportional to $\Omega(\spin)$ scales as $(-1)^{2\spin}$ under the $1\leftrightarrow2$ exchange\footnote{Unlike the AHH minimal coupling, the IR minimal coupling \eqref{emweinbergminimalthreepoint1} exists only for bosonic higher-spins in which case $(-1)^{2\spin}=1$. For fermions, the IR minimal coupling has unequal number of $\SHAA{1^{\bfi}}{2^{\bfj}}$ and $\SHBB{1^{\bfi}}{2^{\bfj}}$.} and has a double pole, the terms proportional to $\Upsilon^{(t)}$ and $\Upsilon^{(u)}$ contributes only at $(t+m^2)$ and $(u+m^2)$ respectively.

Note that even though the opposite helicity amplitude has the similar function form as UV minimal, the same helicity amplitude for the IR minimal doesn't seem to have a form as in the case for UV minimal amplitude.
\subsection{Non-abelian Compton amplitudes}
Similarly, the amplitude factorization can be used to compute the Compton amplitude for IR minimal gluon coupling \eqref{gluoncomp3} which is same as the photon coupling assigned with a matrix $T^{A}_{\mathfrak{ij}}$. The $t$-channel residue \eqref{BCFWtchresidue} is multiplied with a matrix $(T^{A_3}T^{A_4})_{\mathfrak{i}_2\mathfrak{i}_1}=-\text{C}_t$. Due to the non-abelian nature of the matrices $T^A$, the $u$-channel residue must differ by a piece which is proportional to $\iimg f^{A_3A_4B}T^{B}_{\mathfrak{i}_2\mathfrak{i}_1}=\text{C}_s$. So the pole in the $t$-channel residue must be interpreted as a massless gluon exchange from the $s$-channel. Following the same computation as for photons, the $t$-channel residue for gluons is given by \eqref{Tchannelresidue} multiplied by $-\text{C}_t$ and $u+m^2$ replaced by $-s$.
\begin{align}
\left.\widetilde{\mathcal{M}}^{(\text{gluon})}\right|_{t=-m^2}&=\tgem^2 \text{C}_t \left[\frac{m^2\mathfrak{G}_\Omega}{s}\Omega(\spin)-\mathfrak{G}_{\Upsilon_t}\Upsilon^{(t)}(\spin) \right],
\end{align}
 Note that this is same as the residue in \eqref{eq:gluonfinalamp} from the Feynman diagram approach. The $u$ channel residue is again given by the $1\leftrightarrow2$ exchange (under which $\text{C}_t\leftrightarrow -\text{C}_u$)
\begin{align}
	\left.\widetilde{\mathcal{M}}^{(\text{gluon})}\right|_{u=-m^2}&=-\tgem^2 \text{C}_u \left[\frac{m^2\mathfrak{G}_\Omega}{s}\Omega(\spin)-\mathfrak{G}_{\Upsilon_u}\Upsilon^{(u)}(\spin) \right],
\end{align}
Since there is no common term at the $t=-m^2$ and $u=-m^2$ residue, the final answer (same as \eqref{eq:gluonfinalamp}) is simply given by the sum of $t$ and $u$ channel contributions. It manifestly includes the correct residue at the $s=0$ pole given by
\begin{align}
\left.\widetilde{\mathcal{M}}^{\text{(gluon)}}\right|_{s=0}=\tgem^2m^2\left[\frac{\text{C}_t}{t+m^2}-\frac{\text{C}_u}{u+m^2}\right]\mathfrak{G}_\Omega \Omega(\spin)
\end{align}
This is same as the $s$ channel residue obtained in \cite{Arkani-Hamed:2017jhn,Cangemi:2023ysz,Cangemi:2022abk} for the UV minimal Compton amplitudes where $\Omega(\spin)$ is replaced by some other scalars of higher-spin polarizations.
 \section{Conclusion and future directions}
 \label{sec:krsconclusion}
 
 In this work, we have shown a simple way to find contact terms using the on-shell approach, which does not rely on Lagrangian. The principle is simple: in a gauge theory, any physical observable should independent of the choice of gauge. Thus scattering amplitude should also be gauge-invariant. In this work, we considered the spinor helicity formalism, where one needs to choose a reference vector while writing down the polarizations of the massless particles. The choice of reference vector is a choice of gauge and thus any scattering amplitude (and other physical observable) should be independent of the choice of reference vector. 
 
 We considered tree-level four-point scattering amplitude of two charged massive bosonic spinning particles and two photons (i.e. electromagnetic Compton amplitudes). In this case, there is no $s$ channel contribution, and the sum of the $t$ and $u$ channel exchange diagrams is not independent of reference vectors (and thus is not gauge invariant). We could cure this problem by adding a suitable local contact term. We also checked that for the choice, $r_3=4$ and $r_4=3$, the contact term is non-vanishing for the choice of the same helicities and vanishing for the choice of opposite helicities. Similar results have been derived for gluons.
 
 In the case of massless SH variables, BCFW and amplitude factorization\cite{Britto:2004ap, Britto:2005fq} has turned out to be an extremely powerful technique. The authors in \cite{Ochirov:2018uyq, Johansson:2019dnu, Ballav:2020ese, Ballav:2021ahg} also explored BCFW-like recursion relation in the massive SH formalism to compute Compton amplitude. We also verified the results we got for photons and gluons with the amplitude factorization technique. Moreover, we compared the results with the UV minimal amplitudes known in the literature. We showed that the IR and the UV minimal amplitudes have the same pole structure. The only difference is in the higher spin factors in the numerator which arises because the IR and the UV minimal couplings are two independent couplings. Moreover, we also showed that the amplitudes exactly matches with the ones obtained in the Lorentz covariant notation.
 
 The key idea of this paper can be implemented to higher point amplitudes or to gravitational Compton amplitudes. It is known that the three-point amplitude of two massive spinning particles and one massless particle is not unique. There are $2\mathcal{S}+1$ different structures \cite{Arkani-Hamed:2019ymq} (here $\mathcal{S}$ is the spin of the particles). This paper focuses on one of those interactions. However, we found that as long as the numerator for any particular channel takes the form given in \eqref{emweinbergminimal21}, the rest of the analysis holds true. It would be interesting to explore this method for the other three-point amplitudes. As we have discussed, the Compton amplitudes for generic three-point interactions are relevant for various reasons: 1) describing the scattering of general spinning compact objects characterized by the spin-induced multipole moments, 2) EFT understanding of the hadronic resonances and 3) writing down the perturbative rules of EFTs of higher spins as a goal to find a UV complete theory.

 One can also ask similar questions about the gravitons. Just like in the case of Yang-Mills, we need to consider graviton self-coupling and the $s$-channel contribution. For the gravitons, we need to consider the minimal vertex prescribed by the equivalence principle \cite{Weinberg:1964ew}. Once the result is available one can also check whether the result is consistent with double copy \cite{Kawai:1985xq, Bern:2008qj, Bern:2010ue, Cachazo:2013gna, Cachazo:2013hca, Adamo:2022dcm}. In black hole physics, one would also like to compute amplitudes with $2$ massive particles and $n\geq 2$ massless particles. This method could be generalized to a higher loop amplitude as well, and one could check whether it is necessary to add higher point contact terms or if the four-point contact terms would suffice. In the context of gravitational wave classical limit of the scattering amplitudes is being considered \cite{Guevara:2018wpp, Arkani-Hamed:2019ymq, Cangemi:2023ysz}. One could also consider the classical limit of this amplitude \footnote{We are thankful to Arkajyoti Manna for the question and for discussions on this point.}.

 The gravitational Compton amplitudes are extremely useful in the black hole literature \cite{Vines:2017hyw, Buonanno:2022pgc, Chung:2018kqs, Bautista:2021wfy, Bautista:2022wjf}. In the case of electromagnetic interaction, the Rosenbluth formula\footnote{This problem was suggested to us by R Loganayagam.} describes the differential cross-section for the scattering of an electron from a proton due to the exchange of a single photon. One could derive a similar formula for higher-spin particles (instead of protons) as they may be applicable to scattering from highly spinning nuclei.

 \paragraph{Acknowledgement}: We are thankful to R Loganayagam, Manoj Mandal, Julio P. Martinez, Raj Patil \& Manav Shah for many useful discussions. We thank Snehasis Das \& Manoj Mandal for many comments on the first version of the draft. We thank Alok Laddha, Arkajyoti Manna, Shibashis Mukhopadhyay, Raj Patil, Arnab Priya Saha \& Manav Shah for their comment on a preliminary version of the draft. We thank Ishmam Mahbub and both the referees for valuable comments on the last version of the draft. AK gratefully acknowledges the support from IISER Bhopal through the fellowship for PhD students. RS gratefully acknowledges support from CSIR. We thank the members of \href{https://sites.google.com/iiserb.ac.in/iiserbstrings/home}{strings@iiserb}, Department of Physics, IISERB, for providing a vibrant atmosphere. This work was presented in the poster session at the \href{https://iitrpr.ac.in/nsm/}{National strings meeting} at IIT Ropar, at \href{https://www.icts.res.in/program/PosG}{Positive Geometry in Scattering Amplitudes and Cosmological Correlators} at ICTS, and at the \href{https://indico.global/event/13171/}{Amplitudes 2025} conference at SNU, Korea. We thank the organisers for this opportunity. We are thankful to Alfredo Guevara, Andreas Helset, Sangmin Lee, Andres Luna, Alexander Ochirov, Raj Patil, \& Trevor Scheopner for many useful discussions during the Amplitudes 2025 conference. AR would like to thank ICTP for hosting him through the associateship programme towards the completion of the project.

 Finally, we are grateful to the people of India for their generous support for research in basic sciences. 
 
 \vspace{20pt}

 \newpage
 \appendix
 \section{Notation and convention}
 \label{sec:krsnotationandconv}
 Here, we list the notation and conversion followed in this work
 \begin{subequations}   
 	\begin{eqnarray} 
 		\textrm{Spacetime metric}\quad \quad && \eta_{\mu \nu }=\diag(-1,1,\cdots,1)
 		\\
 		\textrm{Index for massless particle/Null vector}\quad \quad &&  i,j
 		\\
 		\textrm{Index for massive particle/Time-like vector}\quad \quad && a,b 
 		\\
 		\textrm{Lorentz indices}\quad \quad && \mu,\nu 
 		\\
 		\textrm{Momentum}\quad \quad && k_\mu 
 		\\
 		\textrm{Mandestam variables}\quad \quad && s,t,u
 		\\
 		\textrm{dotted $SL(2,\mathbb{C})$ indices}\quad \quad && \dot \alpha, \dot \beta
 		\\
 		\textrm{un-dotted $SL(2,\mathbb{C})$ indices} \quad \quad &&\alpha, \beta
 		\\
 		\textrm{massive little group indices}\quad \quad && I, J, \bfi, \bfj 
 		\\
 		\textrm{Reference vectors}\quad \quad && r,\tilde r
 		\\
 		\textrm{Spin}\quad \quad && \mathcal{S} 
 		\\
 		\textrm{Helicity}\quad \quad && h
 		\\
 		\textrm{Polarization of a massless particle}\quad \quad && \epsilon_\mu 
 		\\
 		\textrm{Polarization of a massive particle}\quad \quad && \zeta_\mu 
 		\\
 		\textrm{Single variable Heaviside theta function} \quad\quad && H(x) 
 		\\
 		\textrm{Two-variable Heaviside theta function} \quad\quad && \mathfrak{H}_{x_1,x_2}=H(x_1)H(x_2)
 		\\
 		\textrm{Mass parameters} \quad\quad && m,m_a
 		\\
 		\textrm{Spinor helicity $x$ factor} \quad\quad && x, x_{ia}(h_i)
 		\\
 		\textrm{Coupling contants} \quad\quad && e
 		\\
 		\textrm{Adjoint indices of gauge group} \quad\quad && A,B
 		\\
 		\textrm{Flavour index for matter fields} \quad\quad && \mathfrak{i},  \mathfrak{j}
 	\end{eqnarray} 
 \end{subequations} 
 We follow the following convention for the Mandelstam variables 
 \begin{align}
 	s=(k_1+k_2)^2
 \quad,\quad 
 	t=(k_1+k_4)^2\quad \text{and}
 \quad u=(k_1+k_3)^2
 	\label{hscomp22} 
 \end{align}
 This is the same as the convention in Green-Schwarz-Witten \cite{Green:2012oqa} \footnote{vol.1 page 373, 378.} but different from Polchinski \cite{Polchinski:1998rq}. We also follow the convention such that all the {\it external particles} are {\it incoming}.  
 \subsection{Useful formulae in the SH formalism}
 We are following the convention where particles $1$ and $2$ are massive particles, and $3$ and $4$ are photons.  
 \begin{equation}
 	(k_1)^2=-m^2=(k_2)^2
 	\qquad,\qquad
 	(k_3)^2=0=(k_4)^2
 \end{equation}
 In that convention, here is a list of formulae for 4-point Compton amplitude.
 \begin{eqnarray}
 	&&	\langle 3|2|3]=t+m^2
 	\qquad,\qquad
 	\langle 4|1|4]=t+m^2
 	\nonumber\\
 	&&	\langle 3|1|3]=u+m^2
 	\qquad,\qquad
 	\langle 4|2|4]=u+m^2
 	\nonumber\\		
 	&&
 	\langle 3|4|3]=\langle 34\rangle[34]=s=-t-u-2m^2
 	\nonumber\\	
 	&&
 	\langle 1|2|1]=s+2m^2
 	\qquad,\qquad
 	\langle 2|1|2]=s+2m^2
 	\nonumber\\	
 	&&
 	\langle 3|1|4]\langle 4|1|3]=m^2 s+(u+m^2)(t+m^2)
 	\nonumber\\
 	&&\langle 3|2|4]=-\langle 3|1|4]\qquad,\qquad
 	\langle 4|2|3]=-\langle 4|1|3]
 	\nonumber\\
 	&&
 	\langle 3|2|3]=-\langle 3|1|3]\qquad,\qquad
 	\langle 4|2|4]=-\langle 4|1|4]
 \end{eqnarray}
 
  \subsection{Various Symbolic expressions}
 There are many symbolic expressions which we have used in the article. There are two types of symbolic expressions which are either dependent or independent of the helicities of the external photons/gluons. The helicity independent scalars are as follows
\small
\begin{align}
\Omega(\spin)&=\mathcal{C}^2_\spin \left(\SHAA{1^{\bfi_{j}}}{2^{\bfj_{j}}} \SHBB{1^{\bfi_{j+1}}}{2^{\bfj_{j+1}}}\right)^{\hat{\otimes}_\spin }	\\
\Upsilon^{(t)}(\mathcal{S})&=	\mathcal{C}_{\mathcal{S}}^2\sum_{b=0}^{\mathcal{S}-1}\frac{(-1)^{\mathcal{S}-b}}{m^{2(\mathcal{S}-b)}2^{\mathcal{S}-b}}\tilde B(\mathcal{S},b)
\Big(\langle 1^{\bfi_i}| 4 |1^{\bfi_{i+1}}]\langle 2^{\bfj_i}| 3 |2^{\bfj_{i+1}}] \Big)^{\widehat \otimes_{\mathcal{S}-b-1}}\Big(
\langle 1^{\bfi_j}2^{\bfj_j}\rangle [1^{\bfi_{j+1}}2^{\bfj_{j+1}}]
\Big)^{\widehat \otimes_b},\\
\Upsilon^{(u)}(\mathcal{S})&=	\mathcal{C}_{\mathcal{S}}^2\sum_{b=0}^{\mathcal{S}-1}\frac{(-1)^{\mathcal{S}-b}}{m^{2(\mathcal{S}-b)}2^{\mathcal{S}-b}}\tilde B(\mathcal{S},b)
\Big(\langle 1^{\bfi_i}| 3 |1^{\bfi_{i+1}}]\langle 2^{\bfj_i}| 4 |2^{\bfj_{i+1}}] \Big)^{\widehat \otimes_{\mathcal{S}-b-1}}\Big(
\langle 1^{\bfi_j}2^{\bfj_j}\rangle [1^{\bfi_{j+1}}2^{\bfj_{j+1}}]
\Big)^{\widehat \otimes_b},\\
	{\mathcal{C}_\mathcal{S}}^2&=\frac{1}{(m^2)^{\mathcal{S}}}\quad,\quad \mathfrak{H}_{h_3,h_4}=H(h_3)H(h_4),\\
		\tilde B(\mathcal{S},b) &= \sum\limits_{a=0}^{\lfloor \frac{\mathcal{S}-b}{2} \rfloor}  {}^{\mathcal{S}-2a}\mathrm{C}_b \, \left(-\frac{1}{2}\right)^a\frac{ \mathcal{S}! (2 \mathcal{S}-2 a-1)\text{!!}}{a! (2 \mathcal{S}-1)\text{!!} (\mathcal{S}-2 a)!}
\end{align}
\normalsize
Whereas the helicity dependent scalars are as follows
\small
\begin{align}
		\mathcal{G}_{\Omega}(h_3,h_4)=~&\frac{[34]\langle r_{3}\,r_{4}\rangle}{\langle 3\,r_{3}\rangle\langle 4\,r_{4}\rangle}
	\mathfrak{H}_{h_3,h_4}
	+
	\frac{\langle 4\,r_{3}\rangle[3\,r_{4}]}{\langle 3\,r_{3}\rangle[4\,r_{4}]}
	\mathfrak{H}_{h_3,-h_4}
	+
	\frac{\langle 3\,r_{4}\rangle[ 4\,r_{3}]}{[3\,r_{3}]\langle4\,r_{4} \rangle}
	\mathfrak{H}_{-h_3,h_4}
	+
	\frac{\langle 3\,4\rangle[r_{3}\,r_{4}]}{[3\,r_{3}][4\,r_{4}]}
	\mathfrak{H}_{-h_3,-h_4},\\
		\mathcal{G}_{\Upsilon^{(t)}}(h_3,h_4)=~&-\Big[\frac{\SHAA{1^{\bfi_1}}{r_4}\SHAB{r_3}{2}{3} \SHAB{2^{\bfj_1}}{3}{2^{\bfj_2}}\SHBB{1^{\bfi_2}}{4}}{m^4\SHAA{3}{r_3}\SHAA{4}{r_4}}+
	\frac{\SHAA{2^{\bfj_1}}{r_3}\SHBB{1^{\bfi_1}}{4}\SHBB{1^{\bfi_2}}{4}\SHBB{2^{\bfj_2}}{3}}{m^3\SHAA{3}{r_3}}
	\Big]	\mathfrak{H}_{h_3,h_4}
	\nonumber
	\\
	&+\Big[\frac{\SHBB{1^{\bfi_1}}{r_4}\SHAB{r_3}{2}{3} \SHAB{2^{\bfj_1}}{3}{2^{\bfj_2}}\SHAA{1^{\bfi_2}}{4}}{m^4\SHAA{3}{r_3}\SHBB{4}{r_4}}+
	\frac{\SHAA{2^{\bfj_1}}{r_3}\SHAA{1^{\bfi_1}}{4}\SHAA{1^{\bfi_2}}{4}\SHBB{2^{\bfj_2}}{3}}{m^3\SHAA{3}{r_3}}
	\Big]		\mathfrak{H}_{h_3,-h_4}
	\\
	&+\Big[\frac{\SHAA{1^{\bfi_1}}{r_4}\SHAB{3}{2}{r_3} \SHAB{2^{\bfj_1}}{3}{2^{\bfj_2}}\SHBB{1^{\bfi_2}}{4}}{m^4\SHBB{3}{r_3}\SHAA{4}{r_4}}+
	\frac{\SHBB{2^{\bfj_1}}{r_3}\SHBB{1^{\bfi_1}}{4}\SHBB{1^{\bfi_2}}{4}\SHAA{2^{\bfj_2}}{3}}{m^3\SHBB{3}{r_3}}
	\Big]		\mathfrak{H}_{-h_3,h_4}
	\nonumber
	\\
	&-\Big[\frac{\SHBB{1^{\bfi_1}}{r_4}\SHAB{3}{2}{r_3} \SHAB{2^{\bfj_1}}{3}{2^{\bfj_2}}\SHAA{1^{\bfi_2}}{4}}{m^4\SHBB{3}{r_3}\SHBB{4}{r_4}}+
	\frac{\SHBB{2^{\bfj_1}}{r_3}\SHAA{1^{\bfi_1}}{4}\SHAA{1^{\bfi_2}}{4}\SHAA{2^{\bfj_2}}{3}}{m^3\SHBB{3}{r_3}}
	\Big]		\mathfrak{H}_{-h_3,-h_4}
	\nonumber,\\
	\mathcal{F}^{(t)}(h_3,h_4) =~& 
\frac{\langle r_{3}|2|3]\langle r_{4}|1|4]}{m^2\langle 3\,r_{3}\rangle \langle 4\,r_{4}\rangle }\mathfrak{H}_{h_3,h_4}
-\frac{\langle r_{3}|2|3]\langle 4|1|r_{4}]}{m^2\langle 3\,r_{3}\rangle [4\,r_{4}] }\mathfrak{H}_{h_3,-h_4}
-\frac{\langle 3|2|r_{3}]\langle r_{4}|1|4]}{m^2[3\,r_{3}]\langle 4\,r_{4}\rangle }\mathfrak{H}_{-h_3,h_4}
+
\frac{\langle 3|2|r_{3}]\langle 4|1|r_{4}]}{m^2[3\,r_{3}] [4\,r_{4}] }
\mathfrak{H}_{-h_3,-h_4},\\
\mathcal{F}^{(u)}(h_3,h_4) = ~&
\frac{\langle r_{3}|1|3]\langle r_{4}|2|4]}{m^2\langle 3\,r_{3}\rangle \langle 4\,r_{4}\rangle }\mathfrak{H}_{h_3,h_4}
-\frac{\langle r_{3}|1|3]\langle 4|2|r_{4}]}{m^2\langle 3\,r_{3}\rangle [4\,r_{4}] }\mathfrak{H}_{h_3,-h_4}
-\frac{\langle 3|1|r_{3}]\langle r_{4}|2|4]}{m^2[3\,r_{3}]\langle 4\,r_{4}\rangle }\mathfrak{H}_{-h_3,h_4}
+
\frac{\langle 3|1|r_{3}]\langle 4|2|r_{4}]}{m^2[3\,r_{3}] [4\,r_{4}] }
\mathfrak{H}_{-h_3,-h_4},\\
	\mathcal{F}^{(s)}(h_3,h_4) =~& \mathcal{F}^{(t)}(h_3,h_4)-\mathcal{F}^{(u)}(h_3,h_4)+\frac{t-u}{2m^2}\mathcal{G}_\Omega(h_3,h_4)
\end{align}
\normalsize
The remaining symbolic expressions which appear in the final gauge invariant amplitudes are
\small
\begin{align}
	\mathfrak{G}_\Omega(h_3,h_4) &=-\frac{\SHBB{3}{4}^2}{m^2}\mathfrak{H}_{h_3,h_4}
+\frac{\SHAB{3}{1}{4}^2}{m^4}\mathfrak{H}_{-h_3,h_4}
+\frac{\SHAB{4}{1}{3}^2}{m^4}\mathfrak{H}_{h_3,-h_4}
-\frac{\SHAA{3}{4}^2}{m^2}\mathfrak{H}_{-h_3,-h_4} ,\\
\mathfrak{G}_{\Upsilon_t}(h_3,h_4) &= \frac{\SHBB{1^{\bfi_1}}{4}^2\SHBB{2^{\bfj_1}}{3}^2}{m^4}\mathfrak{H}_{h_3,h_4}
- \frac{\SHBB{1^{\bfi_1}}{4}^2\SHAA{2^{\bfj_1}}{3}^2}{m^4}\mathfrak{H}_{-h_3,h_4}-\frac{\SHAA{1^{\bfi_1}}{4}^2\SHBB{2^{\bfj_1}}{3}^2}{m^4}\mathfrak{H}_{h_3,-h_4}
+\frac{\SHAA{1^{\bfi_1}}{4}^2 \SHAA{2^{\bfj_1}}{3}^2}{m^4} \mathfrak{H}_{-h_3,-h_4}\\
\mathfrak{G}_{\Upsilon_u}(h_3,h_4) &= \frac{\SHBB{2^{\bfj_1}}{4}^2\SHBB{1^{\bfi_1}}{3}^2}{m^4}\mathfrak{H}_{h_3,h_4}
- \frac{\SHBB{2^{\bfj_1}}{4}^2\SHAA{1^{\bfi_1}}{3}^2}{m^4}\mathfrak{H}_{-h_3,h_4}-\frac{\SHAA{2^{\bfj_1}}{4}^2\SHBB{1^{\bfi_1}}{3}^2}{m^4}\mathfrak{H}_{h_3,-h_4}
+\frac{\SHAA{2^{\bfj_1}}{4}^2 \SHAA{1^{\bfi_1}}{3}^2}{m^4} \mathfrak{H}_{-h_3,-h_4},
\end{align}
 \normalsize
 
\subsection{Relation between $\mathfrak{G}_{\Upsilon_t}$ and $\mathfrak{G}_\Omega$}\label{app:GOmegarelation}
At the $t=-m^2$ pole these very differently looking terms are related for any helicity by the expression \eqref{OmgeaUpsilontrelation}. Here we show it for $h_3=h_4=+1$ and for $h_3=-h_4=+1$. The product $(u+m^2)\mathcal{G}_{\Upsilon_t}$ at $t=-m^2,u+m^2=-s$ can be written as
\begin{align}
h_3=h_4=+1\qquad:\qquad\left.(u+m^2)\mathfrak{G}_{\Upsilon_t}\right|_{t=-m^2}&=-\SHAB{3}{4}{3}\frac{\SHBB{1^{\bfi_1}}{4}^2\SHBB{2^{\bfj_1}}{3}^2}{m^4},\label{GOmegatpp}\\
h_3=-h_4=+1\qquad:\qquad\left.(u+m^2)\mathfrak{G}_{\Upsilon_t}\right|_{t=-m^2}&=\SHAB{3}{4}{3}\frac{\SHAA{1^{\bfi_1}}{4}^2\SHBB{2^{\bfj_1}}{3}^2}{m^4}\label{GOmegatpm}
\end{align}
Now we use the following identities which are easy to prove by applying Schouten on the LHS
\begin{align}
	\left.\frac{\langle 4| 1| 3] \left[1^{\bfi} \,4\right]}{m}\right|_{t=-m^2}=-[3 \,4] \left\langle 1^{\bfi} \,4\right\rangle \,\label{simpleid1}\\
	\left.\frac{\langle 4| 1| 3] \left\langle 2^{\bfj} \,3\right\rangle }{m}\right|_{t=-m^2}=-\langle 3 \,4\rangle  \left[2^{\bfj} \,3\right]\label{simpleid2}
\end{align}
Now we can simply take the products of the LHS and RHS of these two identities to produce the following two identities
\begin{align}
	\left.\langle3|4|3] \left[1^{\bfi} \,4\right]\left[2^{\bfj} \,3\right]\right|_{t=-m^2}&=[3 \,4]^2 \left\langle 1^{\bfi} \,4\right\rangle \left\langle 2^{\bfj} \,3\right\rangle \,\\
	\left.\langle3|4|3] \left\langle 1^{\bfi} \,4\right\rangle   \left[2^{\bfj} \,3\right]\right|_{t=-m^2}&=\langle 4| 1| 3]^2 \frac{\left[1^{\bfi} \,4\right]\left\langle 2^{\bfj} \,3\right\rangle}{m^2}
\end{align}
Using these in \eqref{GOmegatpp} and \eqref{GOmegatpm} gives the desired identity 
\begin{align}
	\left.\mathfrak{G}_{\Upsilon_t}(h_3 \,h_4)\right|_{t=-m^2}=\mathfrak{G}_{\Omega}(h_3 \,h_4)\frac{\SHAB{1^{\bfi}}{4}{1^{\bfi}}\SHAB{2^{\bfj}}{3}{2^{\bfj}} }{m^2(u+m^2)}
\end{align}
 \section{Conversion formulae}
 \label{krs:conversionformulae}

 In this appendix, we provide conversion formulae for the readers to convert various expressions involving $SO(3,1)$ vectors and tensors to the corresponding expression using the SH variables. In appendix \ref{sec:krsnotationandconv}, we have already mentioned that 	we use label $i$ for massless particles and $a$ for massive particles
 \begin{enumerate}	
 	
 	\item {\bf Massless momenta:} We begin with massless momenta. 
 	\begin{equation}
 		(k_i)^2=0 \qquad\implies\qquad k_i=-|i\rangle[i|
 		\label{shsetup1}
 	\end{equation}
 	Then, the dot product of two massless momenta is given by
 	\begin{equation}
 		k_i\cdot k_j=\frac{1}{2}\langle ij\rangle[ij]
 		\label{shsetup2}
 	\end{equation}
 	This means the on-shell condition is captured by 
 	\begin{equation}
 		\langle ii\rangle[ii]=0
 		\label{shsetup3}
 	\end{equation}
 	
 	\item {\bf Massive momenta:} We are making an assumption that all massive momenta have the same mass. Since we are currently focussing on Compton amplitudes, this is a sensible assumption to make. A time-like vector is represented as 
 	\begin{equation}
 		(k_a)^2<0\qquad\implies\qquad k_a=-|a_{\bfi} \rangle[a^{\bfi}|
 		\label{shsetup11}
 	\end{equation}
 	In this case, the dot product is given by 
 	\begin{equation}
 		k_a\cdot k_b=\frac{1}{2}\langle a^{\bfi}b^{\bfj}\rangle[a_{\bfi}b_{\bfj}]
 		\label{shsetup12}
 	\end{equation}
 	So, the on-shell condition is given by 
 	\begin{equation}
 		\langle a^{\bfi_1}a^{\bfi_2}\rangle[a_{\bfi_1}a_{\bfi_2}]=-2m^2
 		\label{shsetup13}
 	\end{equation}
 	We choose the normalization of the massive spinors given in \eqref{shbasics11}. 
 	
 	\item {\bf Dot product of a massless momenta and a massive momenta}
 	\begin{equation}
 		k_i\cdot k_a=\frac{1}{2}\langle a^{\bfi}i\rangle[a_{\bfi}i]
 		\label{shsetup16}
 	\end{equation}
 	
 	\item {\bf Massless polarization} 
 	
 	The expression for massless polarization are given in \eqref{shbasics15}. The transversality of the polarization follows from the fact $[ii]=0=\langle ii\rangle$. Now we compute dot product two massless polarization
 	\begin{equation}
 		\varepsilon_i^{(+1)}\cdot  \varepsilon_j^{(+1)}
 		=\frac{\langle r_ir_j \rangle[ij]}{\langle ir_i\rangle\langle jr_j\rangle}
 		\qquad,\qquad 
 		\varepsilon_i^{(-1)}\cdot  \varepsilon_j^{(-1)}
 		=\frac{ [r_ir_j]\langle ij \rangle}{[ir_j][ jr_j]}
 		\qquad,\qquad 
 		\varepsilon_i^{(+1)}\cdot  \varepsilon_j^{(-1)}	
 		=\frac{\langle j r_i \rangle[ i r_j] }{\langle i r_i \rangle[ j r_j] }
 		\label{shsetup23}
 	\end{equation}
 	Now we compute the dot product of a polarization tensor and massless momenta
 	\begin{equation}
 		\varepsilon_i^{(+1)}\cdot k_a
 		=-\frac{1}{\sqrt{2}}\frac{\langle a_{\bfi} r_i\rangle [a^{\bfi}i]}{\langle ir_i\rangle }
 		\qquad,\qquad 	
 		\varepsilon_i^{(-1)}\cdot k_a=
 		\frac{1}{\sqrt{2}}\frac{\langle a_{\bfi}i\rangle [ a^{\bfi} r_i]}{[ir_i]}
 		\label{shsetup24}
 	\end{equation}
 	The formula for dot-product of massless polarization and massive momentum. In order to do that, we begin by defining 
 	\begin{align}
 		x_{ia}=\frac{\langle a_{\bfi} r_i\rangle [a^{\bfi}i]}{m\langle ir_i\rangle }
 		\qquad,\qquad 
 		(x_{ia})^\star=-\frac{\langle a_{\bfi}i\rangle [ a^{\bfi} r_i]}{m[ir_i]}
 		\label{shsetup25}
 	\end{align}
 	Then, the dot product of massive momentum and a massless polarization is given by 
 	\begin{equation}
 		\varepsilon_i^{(+1)}\cdot k_a
 		= -\frac{m}{\sqrt{2}}\, x_{ia}
 		\qquad,\qquad 	
 		\varepsilon_i^{(-1)}\cdot k_a= \frac{m}{\sqrt{2}}\, (x_{ia})^\star 
 		\label{shsetup26}
 	\end{equation}

 	\item {\bf Massive polarization}
 	
 	At last, we discuss massive polarization
 	\begin{equation}
 		\zeta_a^{{\bfi_1\bfi_2}}=\sqrt{2}\frac{|a^{({\bfi_1}}]\langle a^{{\bfi_2})}|}{m}
 		\label{shsetup31}
 	\end{equation}
 	The dot product of one massive momentum and polarization 
 	\begin{equation}
 		\zeta_a\cdot k_b=-\frac{1}{\sqrt{2}m}[a^{({\bfi_1}}b^{{\bfj}}]\langle a^{{\bfi_2})}b_{{\bfj}}\rangle
 		\label{shsetup32}
 	\end{equation}
 	So, the transversality condition simply become 
 	\begin{equation}
 		\zeta_a\cdot k_a=0\qquad \implies\qquad 
 		[a^{({\bfi_1}} a^{\bfi }]
 		\langle a^{\bfi_2)}a_{\bfi }\rangle=0
 		\label{shsetup33}
 	\end{equation}
 	The dot product of one massless momentum and polarization 
 	\begin{equation}
 		\zeta_a\cdot k_i=-\frac{1}{\sqrt{2}m}[a^{({\bfi_1}}i]\langle a^{{\bfi_2})}i\rangle
 		\label{shsetup34}
 	\end{equation}
 	The product of two massive polarizations is given by  
 	\begin{equation}
 		\zeta_a\cdot \zeta_b = -\frac{1}{m^2}[a^{\bfi_1}b^{\bfj_1}]\langle a^{\bfi_2}b^{\bfj_2}\rangle
 		\label{shsetup35}
 	\end{equation}
 	The product of one massless one massive polarization is given by 
 	\begin{equation}
 		\zeta_a\cdot \epsilon_i^{(+1)}=\frac{1}{m}\frac{[r_ia^{(\bfi_1}]\langle a^{\bfi_2)}i\rangle  }{\langle i\,r_i\rangle }
 		\qquad,\qquad
 		\zeta_a\cdot \epsilon_i^{(-1)}=\frac{1}{m}\frac{\langle r_ia^{(\bfi_1}\rangle[ a^{\bfi_2)}i]  }{\langle i\,r_i\rangle }
 		\label{shsetup36}
 	\end{equation}
 \end{enumerate}

 \subsection{Reference vector}
 \label{sec:krsreferencevecappendix}
 
 Polarizations with two different reference vectors differ by a number of times the momentum,
 \begin{align}
 	(\varepsilon^{(+1)}_i(r))^\mu-(\varepsilon^{(+1)}_i(\tilde r ))^\mu=\sqrt{2}k^\mu_i\frac{\braket{r\,\tilde  r}}{\braket{i\,r}\braket{i\,\tilde r}} 
 \end{align}
 Here, we present the proof.
 \begin{equation}
 	(\varepsilon^{(+1)}_i(r))^\mu-(\varepsilon^{(+1)}_i(\tilde r))^\mu
 	=\frac{1}{\sqrt{2}}\bigg[
 	\frac{|i]_{\alpha}\bra{r}_{\dot{\alpha}}(\bar{\sigma}^\mu)^{\dot{\alpha}\alpha}
 	}{\braket{i\,r}}
 	-\frac{|i]_{\alpha}\bra{\tilde r}_{\dot{\alpha}}(\bar{\sigma}^\mu)^{\dot{\alpha}\alpha}
 	}{\braket{i\,\tilde r}}	\bigg]
 \end{equation}
 Taking the common denominator, we obtain 
 \begin{eqnarray}
 	&&\frac{1}{\sqrt{2}}\bigg[
 	\frac{|i]_{\alpha}\bra{r}_{\dot{\alpha}}(\bar{\sigma}^\mu)^{\dot{\alpha}\alpha}
 		\bra{i}_{\dot{\beta}}\bra{\tilde r}_{\dot{\gamma}}\epsilon^{\dot{\beta}\dot{\gamma}}
 		-|i]_{\alpha}\bra{\tilde r}_{\dot{\alpha}}(\bar{\sigma}^\mu)^{\dot{\alpha}\alpha}
 		\bra{i}_{\dot{\beta}}\bra{r}_{\dot{\gamma}}\epsilon^{\dot{\beta}\dot{\gamma}}}{\braket{ir}\braket{is}}
 	\bigg]
 	\\
 	&=&	\frac{1}{\sqrt{2}}\bigg[
 	\bra{r}_{\dot{\alpha}}\bra{\tilde r}_{\dot{\gamma}}-\bra{\tilde r}_{\dot{\alpha}}\bra{r}_{\dot{\gamma}}
 	\bigg]\frac{|i]_{\alpha}\bra{i}_{\dot{\beta}}\epsilon^{\dot{\beta}\dot{\gamma}}(\bar{\sigma}^\mu)^{\dot{\alpha}\alpha}}{\braket{i\,r}\braket{i\tilde r}},
 \end{eqnarray}
 Now, the quantity in the bracket can be simplified using Schouten identity to obtain 
 \begin{equation}
 	\frac{1}{\sqrt{2}}\braket{r\,\tilde r}\epsilon_{\dot{\alpha}\dot{\gamma}}\frac{|i]_{\alpha}\langle i|_{\dot{\beta}}\epsilon^{\dot{\beta}\dot{\gamma}}(\bar{\sigma}^\mu)^{\dot{\alpha}\alpha}}{\braket{i\,r}\braket{i\,\tilde r}}
 	=\sqrt{2}k_i^\mu\frac{\braket{r\,\tilde r}}{\braket{i\,r}\braket{i\,\tilde r}}
 \end{equation}
 This is equivalent to the following statement 
 \begin{eqnarray}
 	\epsilon^\mu(p) \longrightarrow \epsilon^\mu(p) + \alpha\,  p^\mu
 \end{eqnarray}
 
{ \section{Off-Shell terms in the three-point vertices}\label{app:offshellterms}
 As we discussed, the stripped vertices obtained from the \eqref{emweinbergminimalthreepoint1} can have terms which vanish on-shell. In this section, we show that such a term doesn't contribute at the poles of the exchange diagrams. For example in eqn \eqref{StrippedVertices} there can be terms which are proportional to the momenta $k_1$ in the first term and $k_2$ in the second term.
 \begin{align}
		-&
	\iimg \tgem\frac{\mathcal{C}_{\mathcal{S}}}{2^{\mathcal{S}/2}}\,
	x_{32}  \Big(|2^{\bfj_{i}}]_{\alpha}\langle 2^{\bfj_{i+1}}|_{\dot \alpha}+c_1 \frac{\SHAB{2^{\bfj_{i}}}{3}{2^{\bfj_{i+1}}}}{m^2} (k_1+k_4)_{\alpha\dot{\alpha}}\,\Big)^{\widehat \otimes_\mathcal{S} } 
	\quad,\\
	&\iimg \tgem\frac{\mathcal{C}_{\mathcal{S}}}{2^{\mathcal{S}/2}}\,\,
	x_{41}  \Big(|1^{\bfi_{i}}]_{\alpha}\langle 1^{\bfi_{i+1}}|_{\dot \alpha}+c_2 \frac{\SHAB{1^{\bfi_{i}}}{4}{1^{\bfi_{i+1}}}}{m^2} (k_2+k_3)_{\alpha\dot{\alpha}}\,\Big)^{\widehat \otimes_\mathcal{S} }
 \end{align}
 Where $c_1,c_2$ are arbitrary dimensionless factors constructed using the external momentas. The terms proportional to $c_1,c_2$ drops out in the unstripped on-shell coupling because the polarizations are transverse. Using these stripped vertices, the $t$-channel numerator (say for $\spin=1$) will get extra contributions from the terms with a momentum factor on either side of the spin $1$ projector
 \begin{align}
 c_1	(k_1+k_4)_{\alpha_1\dot{\alpha}_1}
 \left(	\Theta^{\alpha_1 \dot\alpha_1 \alpha_2  \dot\alpha_2}(k_1+k_4)\right) |1^{(\bfi_{1}}]_{\alpha_2}\langle 1^{\bfi_{2})}|_{\dot{\alpha}_2} &=-c_1\frac{2 \sqrt{2} (t+m^2) \left\langle 1^{(\bfi_{1}}\right| 4| 1^{\bfi_{2})}]}{m^3},\\
 c_2	|2^{(\bfj_{1}}]_{\alpha_1}\langle 2^{\bfj_{2})}|_{\dot{\alpha}_1} 
 \left(	\Theta^{\alpha_1 \dot\alpha_1 \alpha_2  \dot\alpha_2}(k_1+k_4)\right) (k_2+k_3)_{\alpha_2\dot{\alpha}_2}&=
 	-c_2\frac{2 \sqrt{2} (t+m^2) \left\langle 2^{(\bfj_{1}}\right| 3| 2^{\bfj_{2})}]}{m^3},\\
 c_1c_2	(k_1+k_4)_{\alpha_1\dot{\alpha}_1}
\left( 	\Theta^{\alpha_1 \dot\alpha_1 \alpha_2  \dot\alpha_2}(k_1+k_4)\right)
 	(k_2+k_3)_{\alpha_2\dot{\alpha}_2}&=-c_1c_2\frac{4 (t+m^2) t}{m^2}
 \end{align} 
 All such terms do not contribute at the $t=-m^2$ pole because of the $(t+m^2)$ factor in the numerator. Similar results hold true for $u$-channel which is just $1\leftrightarrow2$ exchange of the $t$-channel. We omit the proof but a more general statement is that ``Two three-point vertices which differ off-shell but are same on-shell gives the same residue at the poles in the exchange diagrams."}
 
 \bibliographystyle{utphys.bst}
 \bibliography{shcomptonbiblio.bib} 
 
 \end{document}